\documentclass[iop,numberedappendix]{emulateapj}

\newcommand{\Dnu}{\mbox{$\Delta \nu$}}
\newcommand{\dnu}[1]{\mbox{$\delta \nu_{#1}$}}
\newcommand{\acena}{\mbox{$\alpha$~Cen~A}}
\newcommand{\acenb}{\mbox{$\alpha$~Cen~B}}

\newcommand{\bhyi}{\mbox{$\beta$~Hyi}}

\newcommand{\cms}{\mbox{cm\,s$^{-1}$}}

\newcommand{\kms}{\mbox{km\,s$^{-1}$}}
\newcommand{\muHz}{\mbox{$\mu$Hz}}

\newcommand{\half}{{\textstyle\frac{1}{2}}}

\slugcomment{accepted by ApJ}

\shorttitle{Oscillations in Procyon. II. Frequencies}
\shortauthors{Bedding et al.}

\begin{document}

\title{
A multi-site campaign to measure solar-like oscillations in Procyon.\\
II. Mode frequencies}

\author{
Timothy~R.~Bedding,\altaffilmark{1} 
Hans~Kjeldsen,\altaffilmark{2} 
Tiago~L.~Campante,\altaffilmark{2,3} 
Thierry~Appourchaux,\altaffilmark{4} 
Alfio~Bonanno,\altaffilmark{5} 
William~J.~Chaplin,\altaffilmark{6} 
Rafael~A.~Garcia,\altaffilmark{7} 
Milena~Marti{\'c},\altaffilmark{8} 
Benoit~Mosser,\altaffilmark{9} 
R.~Paul~Butler,\altaffilmark{10} 
Hans~Bruntt,\altaffilmark{1,11} 
L\'aszl\'o~L.~Kiss,\altaffilmark{1,12} 
Simon~J.~O'Toole,\altaffilmark{13} 
Eiji~Kambe,\altaffilmark{14} 
Hiroyasu~Ando,\altaffilmark{15} 
Hideyuki~Izumiura,\altaffilmark{14} 
Bun'ei~Sato,\altaffilmark{16} 
Michael~Hartmann,\altaffilmark{17} 
Artie~Hatzes,\altaffilmark{17} 
Caroline~Barban,\altaffilmark{11} 
Gabrielle~Berthomieu,\altaffilmark{18} 
Eric~Michel,\altaffilmark{11} 
Janine~Provost,\altaffilmark{18} 
Sylvaine~Turck-Chi\`eze,\altaffilmark{7} 
Jean-Claude~Lebrun,\altaffilmark{8} 
Jerome~Schmitt,\altaffilmark{19} 
Jean-Loup~Bertaux,\altaffilmark{8} 
Serena~Benatti,\altaffilmark{20} 
Riccardo~U.~Claudi,\altaffilmark{21} 
Rosario~Cosentino,\altaffilmark{5} 
Silvio~Leccia,\altaffilmark{22} 
S{\o}ren~Frandsen,\altaffilmark{2} 
Karsten~Brogaard,\altaffilmark{2} 
Lars~Glowienka,\altaffilmark{2} 
Frank~Grundahl,\altaffilmark{2} 
Eric~Stempels,\altaffilmark{23} 
Torben~Arentoft,\altaffilmark{2} 
Micha\"el~Bazot,\altaffilmark{2} 
J{\o}rgen~Christensen-Dalsgaard,\altaffilmark{2} 
Thomas~H.~Dall,\altaffilmark{24} 
Christoffer~Karoff,\altaffilmark{2} 
Jens~Lundgreen-Nielsen,\altaffilmark{2} 
Fabien~Carrier,\altaffilmark{25} 
Patrick~Eggenberger,\altaffilmark{26} 
Danuta~Sosnowska,\altaffilmark{27} 
Robert~A.~Wittenmyer,\altaffilmark{28,29} 
Michael~Endl,\altaffilmark{28} 
Travis~S.~Metcalfe,\altaffilmark{30} 
Saskia~Hekker,\altaffilmark{6,31} and 
Sabine~Reffert\altaffilmark{32} 
}
\altaffiltext{1}{{Sydney Institute for Astronomy (SIfA), School of Physics, University of Sydney, NSW 2006, Australia; bedding@physics.usyd.edu.au}}
\altaffiltext{2}{{Danish AsteroSeismology Centre (DASC), Department of Physics and Astronomy, Aarhus University, DK-8000 Aarhus C, Denmark}}
\altaffiltext{3}{{Centro de Astrof{\'\i}sica da Universidade do Porto, Rua das Estrelas, 4150-762 Porto, Portugal}}
\altaffiltext{4}{{Institut d'Astrophysique Spatiale, Universit{\'e} Paris XI-CNRS, B{\^a}timent 121, 91405 Orsay Cedex, France}}
\altaffiltext{5}{{INAF -- Osservatorio Astrofisico di Catania, via S. Sofia 78, 95123 Catania, Italy}}
\altaffiltext{6}{{School of Physics and Astronomy, University of Birmingham, Edgbaston, Birmingham B15 2TT, UK}}
\altaffiltext{7}{{DAPNIA/DSM/Service d'Astrophysique, CEA/Saclay, 91191 Gif-sur-Yvette Cedex, France}}
\altaffiltext{8}{LATMOS, University of Versailles St Quentin, CNRS, BP 3, 91371 Verrieres le Buisson Cedex, France}
\altaffiltext{9}{{LESIA, CNRS, Universit\'e Pierre et Marie Curie, Universit\'e Denis Diderot, Observatoire de Paris, 92195 Meudon cedex, France}}
\altaffiltext{10}{{Carnegie Institution of Washington, Department of Terrestrial Magnetism, 5241 Broad Branch Road NW, Washington, DC 20015-1305}}
\altaffiltext{11}{Laboratoire AIM, CEA/DSM -- CNRS - Universit\'e Paris Diderot -- IRFU/SAp, 91191 Gif-sur-Yvette Cedex, France}
\altaffiltext{12}{Konkoly Observatory of the Hungarian Academy of Sciences, H-1525 Budapest, P.O. Box 67, Hungary}
\altaffiltext{13}{{Anglo-Australian Observatory, P.O.\,Box 296, Epping, NSW 1710, Australia}}
\altaffiltext{14}{{Okayama Astrophysical Observatory, National Astronomical Observatory of Japan, National Institutes of Natural Sciences, 3037-5 Honjyo, Kamogata, Asakuchi, Okayama 719-0232, Japan}}
\altaffiltext{15}{{National Astronomical Observatory of Japan, National Institutes of Natural Sciences, 2-21-1 Osawa, Mitaka, Tokyo 181-8588, Japan}}
\altaffiltext{16}{{Global Edge Institute, Tokyo Institute of Technology 2-12-1-S6-6, Ookayama, Meguro-ku, Tokyo 152-8550, Japan}}
\altaffiltext{17}{{Th\"uringer Landessternwarte Tautenburg, Sternwarte 5, 07778 Tautenburg, Germany}}
\altaffiltext{18}{{Universit\'e de Nice Sophia-Antipolis, CNRS UMR 6202, Laboratoire Cassiop\'ee, Observatoire de la C\^ote d'Azur, BP 4229, 06304 Nice Cedex, France}}
\altaffiltext{19}{{Observatoire de Haute Provence, 04870 St Michel l'Observatoire, France}}
\altaffiltext{20}{CISAS -- University of Padova, Via Venezia 5, 35131, Padova, Italy}
\altaffiltext{21}{{INAF -- Astronomical Observatory of Padua, Vicolo Osservatorio 5, 35122 Padova, Italy}}
\altaffiltext{22}{{INAF -- Osservatorio Astronomico di Capodimonte, Salita Moiariello 16, 80131 Napoli, Italy}}
\altaffiltext{23}{Department of Physics and Astronomy, Box 516, SE-751 20 Uppsala, Sweden}
\altaffiltext{24}{European Southern Observatory, D-85748 Garching, Germany}
\altaffiltext{25}{{Instituut voor Sterrenkunde, Katholieke Universiteit Leuven, Celestijnenlaan 200 B, 3001 Leuven, Belgium}}
\altaffiltext{26}{Observatoire de Gen\`eve, Universit\'e de Gen\`eve, Ch. des Maillettes 51, CH-1290 Sauverny, Switzerland}
\altaffiltext{27}{{Laboratoire d'astrophysique, EPFL Observatoire CH-1290 Versoix}}
\altaffiltext{28}{{McDonald Observatory, University of Texas at Austin, Austin, TX 78712, USA}}
\altaffiltext{29}{School of Physics, University of New South Wales, NSW 2052, Australia}
\altaffiltext{30}{{High Altitude Observatory, National Centre for Atmospheric Research, Boulder, CO 80307-3000 USA}}
\altaffiltext{31}{{Leiden Observatory, Leiden University, 2300 RA Leiden, The Netherlands}}
\altaffiltext{32}{{ZAH-Landessternwarte, 69117 Heidelberg, Germany}}

\begin{abstract} 
{ We have analyzed data from a multi-site campaign to observe oscillations
in the F5 star Procyon.  The data consist of high-precision velocities that
we obtained over more than three weeks with eleven telescopes.  A new
method for adjusting the data weights allows us to suppress the sidelobes
in the power spectrum.  Stacking the power spectrum in a so-called
\'echelle diagram reveals two clear ridges that we identify with even and
odd values of the angular degree ($l=0$ and 2, and $l=1$ and 3,
respectively).  We interpret a strong, narrow peak at 446\,\muHz\ that lies
close to the $l=1$ ridge as a mode with mixed character.  We show that the
frequencies of the ridge centroids and their separations are useful
{diagnostics} for asteroseismology.  In particular, variations in the large
separation appear to indicate a glitch in the sound-speed profile at an
acoustic depth of $\sim$1000\,s.  We list frequencies for 55 modes
extracted from the data spanning 20 radial orders, {a range comparable
to the best solar data,} which will provide valuable constraints for
theoretical models.  A preliminary comparison with published models shows
that the {offset between observed and calculated frequencies} for the
radial modes is very different for Procyon than for the Sun and other cool
stars.  We find the mean lifetime of the modes in Procyon to be
$1.29^{+0.55}_{-0.49}$\,days, which is significantly shorter than the
2--4\,days seen in the Sun.  }

\end{abstract}

\keywords{stars: individual (Procyon~A) --- stars:~oscillations}

\section{Introduction}

{The success of helioseismology and the promise of asteroseismology
have motivated numerous efforts to measure oscillations in solar-type
stars.  These began with ground-based observations \citep[for recent
reviews see][]{B+K2007c,AChDC2008} and now extend to space-based
photometry, particularly with the {\em CoRoT} and {\em Kepler Missions}
\citep[e.g.,][]{MBA2008,GBChD2010}.}

We have carried out a multi-site spectroscopic campaign to measure
oscillations in the F5 star Procyon~A (HR 2943; HD 61421; HIP 37279).  We
obtained high-precision velocity observations over more than three weeks
with eleven telescopes, with almost continuous coverage for the central ten
days.  In Paper~I \citep{AKB2008PaperI} we described the details of the
observations and data reduction, measured the mean oscillation amplitudes,
gave a crude estimate for the mode lifetime and discussed slow variations
in the velocity curve that we attributed to rotational modulation of active
regions.  In this paper we describe the procedure used to extract the mode
parameters, provide a list of oscillation frequencies, and give an improved
estimate of the mode lifetimes.

\section{Properties of Solar-Like Oscillations} \label{sec.solar-like}

We begin with a brief summary of the relevant properties of solar-like
oscillations (for reviews see, for example,
\citealt{B+G94,B+K2003,ChD2004}).  

To a good approximation, in main-sequence stars the cyclic frequencies of
low-degree p-mode oscillations are regularly spaced, following the
asymptotic relation \citep{Tas80,Gou86}:
\begin{equation}
  \nu_{n,l} \approx \Dnu (n + \half l + \epsilon) - l(l+1) D_0.
        \label{eq.asymptotic}
\end{equation}
Here $n$ (the radial order) and $l$ (the angular degree) are integers,
$\Dnu$ (the large separation) depends on the sound travel time across the
whole star, $D_0$ is sensitive to the sound speed near the core and
$\epsilon$ is sensitive to {the reflection properties of} the surface
layers.  It is conventional to define three so-called small frequency
separations that are sensitive to the sound speed in the core: $\dnu{02}$
is the spacing between adjacent modes with $l=0$ and $l=2$ (for which $n$
will differ by 1); $\dnu{13}$ is the spacing between adjacent modes with
$l=1$ and $l=3$ (ditto); and $\dnu{01}$ is the amount by which $l=1$ modes
are offset from the midpoint of the $l=0$ modes on either
side.\footnote{One can also define an equivalent quantity, $\dnu{10}$, as
the offset of $l=0$ modes from the midpoint between the surrounding $l=1$
modes, {so that $\dnu{10} = \nu_{n, 0} - \half(\nu_{n-1, 1} +
\nu_{n,1})$.}}  {To be explicit, for a given radial order, $n$, these
separations are defined as follows:}
\begin{eqnarray}
  \dnu{02} & = & \nu_{n, 0} - \nu_{n-1,2}  \label{eq.dnu02} \\
  \dnu{01} & = & \half(\nu_{n, 0} + \nu_{n+1,0}) - \nu_{n, 1}
       \label{eq.dnu01} \\ 
  \dnu{13} & = & \nu_{n, 1} - \nu_{n-1,3}. \label{eq.dnu13}
\end{eqnarray}

If the asymptotic relation (equation~\ref{eq.asymptotic}) were to hold
exactly, it would follow that {all of these separations would be
independent of $n$ and that} $\dnu{02} = 6 D_0$, $\dnu{13} = 10 D_0$ and
$\dnu{01} = 2 D_0$.  {In practice,} equation~(\ref{eq.asymptotic}) is only
an approximation.  In the Sun and other stars, theoretical models and
observations show that $\Dnu$, $D_0$ and $\epsilon$ vary somewhat with
frequency, and also with~$l$.  Consequently, the small separations also
vary with frequency.

The mode amplitudes are determined by the excitation and damping, which are
stochastic processes involving near-surface convection.  We typically
observe modes over a range of frequencies, which in Procyon is especially
broad (about 400--1400\,\muHz; see Paper~I).  The observed amplitudes also
depend on $l$ via various projection factors (see Table~1 of
\citealt{KBA2008}).  Note in particular that velocity measurements are much
more sensitive to modes with $l=3$ than are intensity measurements.  The
mean mode amplitudes are modified for a given observing run by the
stochastic nature of the excitation, resulting in considerable scatter of
the peak heights about the envelope.

Oscillations in the Sun are long-lived compared to their periods, which
allows their frequencies to be measured very precisely.  However, the
lifetime is not infinite and the damping results in each mode in the power
spectrum being split into multiple peaks under a Lorentzian profile.  The
FWHM of this Lorentzian, which is referred to as the linewidth $\Gamma$, is
inversely proportional to the mode lifetime: $\Gamma = 1/(\pi\tau)$.
{We follow the usual definition that $\tau$ is the time for the mode
amplitude to decay by a factor of~$e$.}  The solar value of $\tau$ for the
strongest modes ranges from 2 to 4\,days, as a decreasing function of
frequency \citep[e.g.,][]{CEI97}.

Procyon is an evolved star, with theoretical models showing that it is
close to, or just past, the end of the main sequence
\citep[e.g.,][]{G+D93,BMM99,CDG99,DiM+ChD2001,KTM2004,ECB2005,PBM2006,BKP2007,GKG2008}.
As such, its oscillation spectrum may show deviations from the regular
comb-like structure described by equation~(\ref{eq.asymptotic}), especially
at low frequencies.  This is because some modes, particularly those with
$l=1$, are shifted by avoided crossings with gravity modes in the stellar
core (also called `mode bumping'; see \citealt{Osa75,ASW77}).  These
so-called `mixed modes' have p-mode character near the surface but g-mode
character in the deep interior.  Some of the theoretical models of Procyon
cited above indeed predict these mixed modes, depending on the evolutionary
state of the star, and we must keep this in mind when attempting to
identify oscillation modes in the power spectrum.  The mixed modes are rich
in information because they probe the stellar core and are very sensitive
to age, but they complicate the task of mode identification.

We should also keep in mind that mixed modes are expected to have
{lower amplitudes and} longer lifetimes (smaller linewidths) than
regular p modes because they have larger mode inertias
\citep[e.g.,][]{ChD2004}.  {Hence, for a data series that is many times
longer than the lifetime of the pure p modes, a mixed mode may appear in
the power spectrum as a narrow peak that is higher than the others, even
though its power (amplitude squared) is not especially large.  }

Another potential complication is that stellar rotation causes modes with
$l \ge 1$ to split into multiplets.  The peaks of these multiplets are
characterized by the azimuthal degree~$m$, which takes on values of~$m = 0,
\pm1, \ldots, \pm l$, with a separation that directly measures the rotation
rate averaged over the region of the star that is sampled by the mode.  The
measurements are particularly difficult because a long time series is
needed to resolve the rotational splittings.  We argue in
Appendix~\ref{app.rotation} that the low value of $v\sin i$ observed in
Procyon implies that rotational splitting of frequencies is not measurable
in our observations.

\section{Weighting the time series} \label{sec.weights}

The time series of velocity observations was obtained over 25 days using 11
telescopes at eight observatories and contains just over 22\,500 data
points.  As discussed in Paper~I, the velocity curve shows slow variations
that we attribute to a combination of instrumental drifts and rotational
modulation of stellar active regions.  We have removed these slow
variations by subtracting all the power below 280$\mu$Hz, to prevent spectral
leakage into higher frequencies that would degrade the oscillation
spectrum.  We take this high-pass-filtered time series of velocities,
together with their associated measurement uncertainties, as the starting
point in our analysis.

\subsection{Noise-optimized weights}

Using weights when analyzing ground-based observations of stellar
oscillations \citep[e.g.,][]{GBK93,FJK95} allows one to take into account
the significant variations in data quality during a typical observing
campaign, especially when two or more telescopes are used.  The usual
practice, which we followed in Paper~I, is to calculate the weights for a
time series from the measurement uncertainties, $\sigma_i$, according to
$w_i=1/\sigma_i^2$.

These ``raw'' weights can then be adjusted to minimize the noise level in
the final power spectrum by identifying and revising those uncertainties
that are too optimistic, and at the same time rescaling the uncertainties
to be in agreement with the actual noise levels in the data.  This
procedure is described in Paper~I and references therein.  The time series
of these noise-optimized weights is shown in Figure~\ref{fig.weights}{\em
a}.  These are the same as those shown in Figure~1{\em d} of Paper~I, but
this time as weights rather than uncertainties.

The power spectrum of Procyon based on these noise-optimized weights is
shown in Figure~\ref{fig.power}{\em a}.  This is the same as shown in
Paper~I (lower panel of Figure~6), except that the power at low
frequencies, which arises from the slow variations, has been removed.  As
described in Paper~I, the noise level above 3\,mHz in this noise-optimized
spectrum is 1.9\,\cms{} in amplitude.  This includes some degree of
spectral leakage from the oscillations and if we high-pass filter the
spectrum up to 3\,mHz to remove the oscillation signal, the noise level
drops to 1.5\,\cms{} in amplitude.

The task of extracting oscillation frequencies from the power spectrum is
complicated by the presence of structure in the spectral window, which are
caused by gaps or otherwise uneven coverage in the time series.  The
spectral window using the noise-optimized weights is shown in
Figure~\ref{fig.window}{\em a}.  Prominent sidelobes at $\pm 11.57\,\muHz$
correspond to aliasing at one cycle per day.  Indeed, the prospect of
reducing these sidelobes is the main reason for acquiring multi-site
observations.  However, even with good coverage the velocity precision
varies greatly, both for a given telescope during the run and from one
telescope to another (see Figure~\ref{fig.weights}{\em a}).  As pointed out
in Paper~I, using these measurement uncertainties as weights has the effect
of increasing the sidelobes in the spectral window.  We now discuss a
technique for addressing this issue.

\subsection{Sidelobe-optimized weights}

Adjusting the weights allows one to suppress the sidelobe structure; the
trade-off is an increase in the noise level.  This technique is routinely
used in radio astronomy when synthesising images from interferometers
\citep[e.g.,][]{H+B74}.  {An extreme case is to set all weights to be
equal, which is the same as not using weights at all.  This is certainly
not optimal because it produces a power spectrum with greatly increased
noise (by a factor of 2.3) but still having significant sidelobes, as can
be seen in Figure 6{\it a} of Paper~I.}  Adjusting the weights on a
night-by-night basis in order to minimize the sidelobes was used in the
analysis of dual-site observations of \acena{} \citep{BKB2004}, \acenb{}
\citep{KBB2005}, and \bhyi{} \citep{BKA2007}.  For our multi-site Procyon
data this is impractical {because of the large number of (partly
overlapping) telescope nights.}  We have developed a more general algorithm
for adjusting weights to minimize the sidelobes (H. Kjeldsen et al., in
prep.).  {The new method, which is superior because it does not assume the
oscillations are coherent over the whole observing run, is based on the
principle that equal weight is given to all segments of the time series.}
The method produces the cleanest possible spectral window in terms of
suppressing the sidelobes, and we have tested it with good results using
published data for $\alpha$~Cen A and B, and $\beta$~Hyi \citep{AKB2010}.

The new method operates with two timescales, $T_1$ and $T_2$.  All data
segments of length $T_1$ (=2\,hr, in this case) are required to have the
same total weight throughout the time series, with the relaxing condition
that variations on time scales longer than $T_2$ (=12\,hr) are retained.
To be explicit, the algorithm works as follows.  We adjust the weights so
that all segments of length $T_1$ have the same total weight.  That is, for
each point $w_i$ in the time series of weights, define $\{S_i\}$ to be the
set of weights in a segment of width $T_1$ centered at that time stamp, and
divide each $w_i$ by the sum of the weights in~$\{S_i\}$.  However, this
adjustment suffers from edge effects, since it gives undue weight to points
adjacent to a gap.  To compensate, we also divide by an asymmetry factor
\begin{equation}
  R = 1+ \left|\frac{\Sigma_{\rm left} - \Sigma_{\rm right}}{\Sigma_{\rm left} +
  \Sigma_{\rm right}}\right|. 
\end{equation}
Here, $\Sigma_{\rm left}$ is the sum of the weights in the segment
$\{S_i\}$ that have time stamps less than $t_i$, and $\Sigma_{\rm right}$
is the sum of the weights in the segment $\{S_i\}$ that have time stamps
greater than $t_i$.  Note that $R$ ranges from 1, for points that are
symmetrically placed in their $T$ bin, up to 2 for points at one edge of
a gap.  

Once the above procedure is done for $T_1$, which is the shortest timescale
on which we wish to adjust the weights, we do it again with $T_2$, which is
the longest timescale for adjusting the weights.  Finally, we divide the
first set of adjusted weights by the second set, and this gives the weights
that we adopt (Figure~\ref{fig.weights}{\em b}).

\subsection{The sidelobe-optimized power spectrum}

Figure~\ref{fig.power}{\em b} shows the power spectrum of Procyon based on
the sidelobe-optimized weights.  The spectral window has improved
tremendously (Figure~\ref{fig.window}{\em b}), while the noise level at
high frequencies (above 3\,mHz) has increased by a factor of 2.0.

The power spectrum now clearly shows a regular series of peaks, which are
even more obvious after smoothing (Figure~\ref{fig.power}{\em c}).  We see
that the large separation of the star is about 55\,\muHz, confirming the
value indicated by several previous studies
\citep{MMM98,MSL99,MLA2004,ECB2004,R+RC2005,LKB2007,GKG2008}.  The very
strong peak at 446\,\muHz{} appears to be a candidate for a mixed mode,
especially given its narrowness {(see Section~\ref{sec.solar-like})}.

Plotting the power spectrum in \'echelle format using a large separation of
56\,\muHz\ (Figure~\ref{fig.echelle.image.WIN56}) clearly shows two ridges,
as expected.\footnote{When making an \'echelle diagram, it is common to
plot $\nu$ versus ($\nu \bmod \Dnu$), in which case each order slopes
upwards slightly.  However, for gray-scale images it is preferable to keep
the orders horizontal, {as was done in the original presentation of the
diagram \citep{GFP83}.}  We have followed that approach in this paper, and
the value given on the vertical axis indicates the frequency at the middle
of each order.}  The upper parts are vertical but the lower parts are
tilted, indicating a change in the large separation as a function of
frequency.  This large amount of curvature in the \'echelle diagram goes a
long way towards explaining the lack of agreement between previous studies
on the mode frequencies of Procyon {(see the list of references given
in the previous paragraph)}.

The advantage of using the sidelobe-optimized weights is demonstrated by
Figure~\ref{fig.echelle.image.SNR56}.  This is the same as
Figure~\ref{fig.echelle.image.WIN56} but for the noise-optimized weights
and the ridges are no longer clearly defined.

\section{Identification of the ridges}
\label{sec.ridge.id}

We know from asymptotic theory (see equation~\ref{eq.asymptotic}) that one
of the ridges in the \'echelle diagram
(Figure~\ref{fig.echelle.image.WIN56}) corresponds to modes with even
degree ($l=0$ and 2) and the other to modes with odd degree ($l=1$ and 3).
However, it is not immediately obvious which is which.  We also need to
keep in mind that the asymptotic relation in evolved stars does not hold
exactly.  We designate the two possibilities Scenario~A, in which the
left-hand ridge in Figure~\ref{fig.echelle.image.WIN56} corresponds to
modes with odd degree, and Scenario~B, in which the same ridge corresponds
to modes with even degree.  Figure~\ref{fig.collapse3} shows the Procyon
power spectrum collapsed along several orders.  We see now double peaks
that suggest the identifications shown, which corresponds to Scenario~B.

We can check that the small separation $\dnu{01}$ has the expected sign.
According to asymptotic theory (equation~\ref{eq.asymptotic}), each $l=1$
mode should be at a slightly lower frequency than the mid-point of the
adjacent $l=0$ modes.  This is indeed the case for the identifications
given in Figure~\ref{fig.collapse3}, but would not be if the even and odd
degrees were reversed.  We should be careful, however, since \dnu{01} has
been observed to have the opposite sign in red giant stars
\citep{CDRB2010,BHS2010}.  

The problem of ridge identification in F stars was first encountered by
\citet{AMA2008} when analysing CoRoT observations of HD~49933 and has been
followed up by numerous authors
\citep{BAB2009,BBC2009,GKW2009,M+A2009,Rox2009,KGG2010}.  Two other F stars
observed by CoRoT have presented the same problem, namely HD~181906
\citep{GRS2009} and HD~181420 \citep{BDB2009}.  A discussion of the issue
was recently given by \citet{B+K2010}, who proposed a solution to the
problem that involves comparing two (or more) stars on a single \'echelle
diagram after first scaling their frequencies.

Figure~\ref{fig.echelle.corot} shows the \'echelle diagram for Procyon
overlaid with scaled frequencies for two stars observed by CoRoT, using the
method described by \citet{B+K2010}.  The filled symbols are scaled
oscillation frequencies for the G0 star HD~49385 observed by CoRoT
\citep{DBM2010}.  The scaling involved multiplying all frequencies by a
factor of 0.993 before plotting them, with this factor being chosen to
align the symbols as closely as possible with the Procyon ridges.  For this
star the CoRoT data gave an unambiguous mode identification, which is
indicated by the symbol shapes.  This confirms that the left-hand ridge of
Procyon corresponds to modes with even $l$ (Scenario~B).  

The open symbols in Figure~\ref{fig.echelle.corot} are oscillation
frequencies for HD~49933 from the revised identification by
\citet[][Scenario~B]{BBC2009}, after multiplying by a scaling factor of
0.6565.  The alignment with HD~49385 was already demonstrated by
\citet{B+K2010}.  We show HD~49933 here for comparison and to draw
attention to the different amounts of bending at the bottom of the
right-hand ($l=1$) ridge for the three stars.  {The CoRoT target that
is most similar to Procyon is HD170987 but unfortunately the $S/N$ ratio is
too low to provide a clear identification of the ridges \citep{MGC2010}.}

The above considerations give us confidence that Scenario~B in Procyon is
the correct identification, and we now proceed on that basis.


\section{Frequencies of the Ridge Centroids}
\label{sec.ridge.centroids}

Our next step in the analysis was to measure the centroids of the two
ridges in the \'echelle diagram.  We first removed the strong peak at
446\,\muHz\ (it was replaced by the mean noise level).  We believe this to
be a mixed mode and its extreme power means that it would significantly distort
the result.  We then smoothed the power spectrum to a resolution of
10\,\muHz\ (FWHM).  To further improve the visibility of the ridges, we
also averaged across several orders, which corresponds to smoothing in the
vertical direction in the \'echelle diagram
\citep{BKB2004,KBB2005,Kar2007}.  That is, for a given value of \Dnu\ we
define the ``order-averaged'' power-spectrum to be
\begin{equation}
    {\rm OAPS}(\nu, \Dnu) = \sum_{j=-4}^4  c_j PS(\nu + j \Dnu).
    \label{eq.OAPS}
\end{equation}
The coefficients $c_j$ are chosen to give a smoothing with a FWHM of
$k\Dnu$:
\begin{equation}
    c_j = c_{-j} = \frac{1}{1 + (2j/k)^2}.
\end{equation}
We show in Figure~\ref{fig.idl9} the OAPS based on smoothing over 4 orders
($k=4.0$), and so we used $(c_0,\ldots, c_4) = (1, 0.8, 0.5, 0.31, 0.2)$.
The OAPS is plotted for three values of the large separations (54, 55 and
56\,\muHz) and they are superimposed.  The three curves are hardly
distinguishable and we see that the positions of the maxima are not
sensitive to the value of \Dnu.

We next calculated a modified version of the OAPS in which the value at
each frequency is the maximum value of the OAPS over a range of large
separations (53--57\,\muHz).  This is basically the same as the comb
response, as used previously by several authors
\citep{KBV95,MMM98,MSL99,LKB2007}.  The maxima of this function define the
centroids of the two ridges, which are shown in
Figure~\ref{fig.echelle.ridges56}.

In Figure~\ref{fig.collapse.ridges} we show the full power spectrum of
Procyon (using sidelobe-optimized weights) collapsed along the ridges.
This is similar to Figure~\ref{fig.collapse3} except that {each order
was shifted before the summation, so as to align the ridge peaks (symbols
in Figure~\ref{fig.echelle.ridges56}) and hence remove the curvature.}
This was done separately for both the even- and odd-degree ridges, as shown
in the two panels of Figure~\ref{fig.collapse.ridges}.  The collapsed
spectrum clearly shows the power corresponding to $l=0$--3, as well as the
extra power from the mixed modes (for this figure, the peak at 446\,\muHz\
has not been removed).

In Section~\ref{sec.clean} below, we use the ridges to guide our
identification of the individual modes.  First, however, we show that some
asteroseismological inferences can be made solely from the ridges
themselves.  {This is explained in more detail in
Appendix~\ref{app.ridges}.}

\subsection{Large separation of the ridges}

Figure~\ref{fig.seps.ridges}{\it a} shows the variation with frequency of
the large separation for each of the two ridges (diamonds and triangles).
{The smoothing across orders (equation~\ref{eq.OAPS}) means that the
ridge frequencies are correlated from one order to the next and so we used
simulations to estimate uncertainties for the ridge centroids.}

The oscillatory behavior of \Dnu\ as a function of frequency seen in
Figure~\ref{fig.seps.ridges}{\it a} is presumably a signature of the helium
ionization zone {\citep[e.g.][]{Gou90}}.  
The oscillation is also seen in Figure~\ref{fig.seps.ridges}{\it b}, which
shows the second differences for the two ridges, defined as follows
\citep[see][]{Gou90,BTCG2004,H+G2007}:
\begin{eqnarray}
  \Delta_2\nu_{n, \rm even} & = & \nu_{n-1,\rm even} - 2 \nu_{n,\rm even} + \nu_{n+1, \rm even}\\
  \Delta_2\nu_{n, \rm  odd} & = & \nu_{n-1,\rm  odd} - 2 \nu_{n,\rm  odd} + \nu_{n+1, \rm  odd}.
\end{eqnarray}
The period of the oscillation is $\sim$500\,\muHz, which implies a glitch
at an acoustic depth that is approximately twice the inverse of this value
\citep{Gou90,H+G2007}, namely $\sim$1000\,s.  {To determine this more
precisely, we calculated the power spectrum of the second differences for
both the odd and even ridges, and measured the highest peak.  We found the
period of the oscillation in the second differences to be
$508\pm18\,\muHz$.}  Comparing this result with theoretical models will be
the subject of a future paper.

{The dotted lines in Figure~\ref{fig.seps.ridges}{\it a} show the variation
of \Dnu\ with frequency calculated from the autocorrelation of the time
series using the method of \citet[][see also \citealt{R+V2006}]{M+A2009}.
The mixed mode at 446\,\muHz\ was first removed and the smoothing filter
had FWHM equal to 3 times the mean large separation.  We see general
agreement with the values calculated from the ridge separations.  Some of
the differences presumably arise because the autocorrelation analysis of
the time series averages the large separation over all degrees.

\subsection{Small separation of the ridges}
\label{sec.ridge.dnu}

Using only the centroids of the ridges, we can measure a small separation
that is useful for asteroseismology.  By analogy with \dnu{01}\ (see
equation~\ref{eq.dnu01}), we define it as the amount by which the odd
ridge is offset from the midpoint of the two adjacent even ridges, with a
positive value corresponding to a leftwards shift (as observed in the Sun).
That is,
\begin{equation}
  \delta\nu_{\rm even,odd} = 
         \frac{\nu_{n,\rm even} + \nu_{n+1,\rm even}}{2} - \nu_{n, \rm odd}.
	 \label{eq.dnu_even_odd}
\end{equation}
Figure~\ref{fig.seps.ridges}{\it c} shows our measurements of this small
separation.\footnote{We could also define a small separation \dnu{\rm
odd,even} to be the amount by which the centroid of the {\rm even} ridge is
offset rightwards from the midpoint of the adjacent {\rm odd} ridges.  This
gives similar results.}  It is related in a simple way to the conventional
small separations \dnu{01}, \dnu{02}, and \dnu{13} (see
Appendix~\ref{app.ridges} for details) and so, like them, it gives
information about the sound speed in the core.  Our measurements of this
small separation can be compared with theoretical models using the
equations in Appendix~\ref{app.ridges} \citep[e.g., see][]{ChD+H2009}.

\section{Frequencies of individual modes}
\label{sec.clean}

We have extracted oscillation frequencies from the time series using the
standard procedure of iterative sine-wave fitting.  Each step of the
iteration involves finding the strongest peak in the sidelobe-optimized
power spectrum and subtracting the corresponding sinusoid from the time
series.  Figure~\ref{fig.echelle.clean.sidelobe} shows the result.  The two
ridges are clearly visible but the finite mode lifetime causes many modes
to be split into two or more peaks.  We might also be tempted to propose
that some of the multiple peaks are due to rotational splitting but, as
shown in Appendix~\ref{app.rotation}, this is unlikely to be the case.

Deciding on a final list of mode frequencies with correct $l$
identifications is somewhat subjective.  To guide this process, we used the
ridge centroids shown in Figure~\ref{fig.echelle.ridges56} as well as the
small separations $\dnu{02}$ and $\dnu{13}$ from the collapsed power
spectrum (see Figures~\ref{fig.collapse3} and~\ref{fig.collapse.ridges}).
Each frequency extracted using iterative sine-wave fitting that lay close
to a ridge was assigned an $l$ value and multiple peaks from the same mode
were averaged.  The final mode frequencies are listed in
Table~\ref{tab.freq.matrix}, while peaks with $S/N \ge 3.5$ that we have not
identified are listed in Table~\ref{tab.freq.list.other}.
Figures~\ref{fig.power.zoom} and~\ref{fig.echelle.cleanid} show these peaks
overlaid on the sidelobe-optimized power spectrum.
{Figure~\ref{fig.seps.freq} shows the three small separations
(equations~\ref{eq.dnu02}--\ref{eq.dnu13}) as calculated from the
frequencies listed in Table~\ref{tab.freq.matrix}. }
{The uncertainties in the mode frequencies are shown in parentheses in
Table~\ref{tab.freq.matrix}.  These depend on the $S/N$ ratio of the peak
and were calibrated using simulations \citep[e.g., see][]{BKA2007}.}

The entries in Table~\ref{tab.freq.list.other} {are mostly false peaks due
to noise and to residuals from the iterative sine-wave fitting,} but may
include some genuine modes.  To check whether some of them may be daily
aliases of each other or of genuine modes, we calculated the differences of
all combinations of frequencies in Tables~\ref{tab.freq.matrix}
and~\ref{tab.freq.list.other}.  The histogram of these pairwise differences
was flat in the vicinity of 11.6\,\muHz\ and showed no excess, confirming
that daily aliases do not contribute significantly to the list of
frequencies in the tables.

{We also checked whether the number peaks in
Table~\ref{tab.freq.list.other} agrees with expectations.  We did this by
analysing a simulated time series that matched the observations in terms of
oscillations properties (frequencies, amplitudes and mode lifetimes), noise
level, window function and distribution of weights.  We extracted peaks
from the simulated power spectrum using iterative sine-wave fitting, as
before, and found the number of ``extra'' peaks (not coinciding with the
oscillation ridges) to be similar to that seen in
Figure~\ref{fig.echelle.clean.sidelobe}.  Finally, we remark that the peak
at 408\,\muHz\ is a candidate for a mixed mode with $l=1$, given that it
lies in the same order as the previously identified mixed mode at
446\,\muHz\ (note that we expect one extra $l=1$ mode to occur at an
avoided crossing).  }

The modes listed in Table~\ref{tab.freq.matrix} span 20 radial orders and
more than a factor of 4 in frequency.  This range is similar to that
obtained from long-term studies of the Sun \citep[e.g.,][]{BCD2009} and is
unprecedented in asteroseismology.  It was made possible by the unusually
broad range of excited modes in Procyon and the high S/N of our data.
Since the stellar background at low frequencies in intensity measurements
is expected to be much higher than for velocity measurements, it seems
unlikely that even the best data from the {\em Kepler Mission} will return
such a wide range of frequencies in a single target.


\section{Mode lifetimes} \label{sec.lifetimes}

As discussed in Section~\ref{sec.solar-like}, if the time series is
sufficiently long then damping causes each mode in the power spectrum to be
split into a series of peaks under a Lorentzian envelope having FWHM
$\Gamma = 1/(\pi\tau)$, where $\tau$ is the mode lifetime.  Our
observations of Procyon are not long enough to resolve the modes into clear
Lorentzians, and instead we see each mode as a small number of peaks
(sometimes one).  Furthermore, the centroid of these peaks may be offset
from the position of the true mode, as illustrated in Figure~1 of
\citet{ADJ90}.  This last feature allows one to use the scatter of the
extracted frequencies about smooth ridges in the \'echelle diagram,
calibrated using simulations, to estimate the mode lifetime
\citep{KBB2005,BKA2007}.  That method cannot be applied to Procyon because
the $l=0$ and $l=2$ ridges are not well-resolved and the $l=1$ ridge is
affected by mixed modes.

Rather than looking at frequency shifts, we have estimated the mode
lifetime from the variations in mode amplitudes (again calibrated using
simulations).  This method is less precise but has the advantage of being
independent of the mode identifications
\citep[e.g.,][]{LKB2007,CKB2007,BKA2007}.  In Paper~I we calculated the
smoothed amplitude curve for Procyon in ten 2-day segments and used the
fluctuations about the mean to make a rough estimate of the mode lifetime:
$\tau = 1.5_{-0.8}^{+1.9}$\,days.  We have attempted to improve on that
estimate by considering the amplitude fluctuations of individual modes, as
has been done for the Sun \citep[e.g.,][]{T+F92,BGG96,C+G98}, but were not
able to produce well-calibrated results for Procyon.

Instead, we have measured the ``peakiness'' of the power spectrum
\citep[see][]{BKA2007} by calculating the ratio between the square of the
mean amplitude of the 15 highest peaks in the range 500--1300\,\muHz\
(found by iterative sine-wave fitting) and the mean power in the same
frequency range.  The value for this ratio from our observations of Procyon
is 6.9.  We made a large number of simulations (3600) having a range of
mode lifetimes and with the observed frequency spectrum, noise level,
window function and weights.  Comparing the simulations with the
observations led to a mode lifetime for Procyon of
$1.29^{+0.55}_{-0.49}$\,days.

This agrees with the value found in Paper~I but is more precise, confirming
that modes in Procyon are significantly more short-lived than those of the
Sun.  As discussed in Section~\ref{sec.solar-like}, the dominant modes in
the Sun have lifetimes of 2--4\,days \citep[e.g.,][]{CEI97}.  The tendency
for hotter stars to have shorter mode lifetimes has recently been discussed
by \citet{CHK2009}.

\section{Fitting to the power spectrum}
\label{sec.fit}

Extracting mode parameters by fitting directly to the power spectrum is
widely used in helioseismology, where the time series extends continuously
for months or even years, and so the individual modes are well-resolved
\citep[e.g.,][]{ADJ90}.  Mode fitting has not been applied to ground-based
observations of solar-type oscillations because these data typically have
shorter durations and significant gaps.  Global fitting has been carried
out on spacecraft data, beginning with the 50-d time series of \acena\
taken with the WIRE spacecraft \citep{FCE2006} and the 60-d light curve of
HD\,49933 from CoRoT \citep{AMA2008}.  Our observations of Procyon are much
shorter than either of these cases but, given the quality of the data and
the spectral window, we considered it worthwhile to attempt a fit.

{Global fits to the Procyon power spectrum were made by several of us.
Here, we present results from} a fit using a Bayesian approach
\citep[e.g.,][]{Gre2005}, which allowed us to include in a straightforward
way our prior knowledge of the oscillation properties.  The parameters to
be extracted were the frequencies, heights and linewidths of the modes.  To
obtain the marginal probability distributions of these parameters and their
associated uncertainties, we employed an APT MCMC (Automated Parallel
Tempering Markov Chain Monte Carlo) algorithm.  It implements the
Metropolis-Hastings sampler by performing a random walk in parameter space
while drawing samples from the posterior distribution \citep{Gre2005}.
Further details of our implementation of the algorithm will be given
elsewhere (T.L. Campante et al., in prep.).

The details of the fitting are as follows:
\begin{itemize}

\item The fitting was performed over 17 orders (5--21) using the
  sidelobe-optimized power spectrum.  In each order we fitted modes with
  $l=0$, 1, and 2, with each individual profile being described by a
  symmetric Lorentzian with FWHM~$\Gamma$ and height~$H$.  The mode
  frequencies were constrained to lie close to the ridges and to have only
  small jumps from one order to the next (a Gaussian prior with $\sigma =
  3\,\muHz$).  {The S/N ratios of modes with $l=3$ were too low to
  permit a fit.  In order to take their power into account, we included
  them in the model with their frequencies fixed by the asymptotic relation
  (equation~\ref{eq.asymptotic}).}

\item The data are not good enough to provide a useful estimate of the
  linewidth of every mode, or even of every order.  Therefore, the
  linewidth was parametrized as a linear function of frequency, defined by
  two parameters $\Gamma_{600}$ and $\Gamma_{1200}$, which are the values
  at 600 and 1200\,\muHz.  These parameters were determined by the fit, in
  which both were assigned a uniform prior in the range 0--10\,\muHz.

 \item The height of each mode is related to the linewidth and amplitude
   according to \citep{CHE2005}:
   \begin{equation}
     H = \frac{2A^2}{\pi \Gamma}.
   \end{equation}
  The amplitudes $A$ of the modes were determined as follows.  For the
  radial modes ($l=0$) we used the smoothed amplitude curve measured from
  our observations, as shown in Figure~10 of Paper~I.  The amplitudes of
  the non-radial modes ($l=1$--3) were then calculated from the radial
  modes using the ratios given in Table~1 of \citet{KBA2008}, namely
  $S_0:S_1:S_2:S_3 = 1.00:1.35:1.02:0.47$.

 \item The background was fitted as a flat function.

 \item We calculated the rotationally-split profiles of the non-radial
 modes using the description given by \citet{G+S2003}.  The inclination
 angle of the rotation axis was fixed at $31^{\circ}$, which is the
 inclination of the binary orbit \citep{GWL2000} and, as discussed in
 Paper~I (Section~4.1), is consistent with the rotational modulation of the
 velocity curve.  The rotational splitting was fixed at 0.7\,\muHz, which
 was chosen to match the observed value of $v\sin i = 3.16$\,\kms\
 \citep{APAL2002}, given the known radius of the star.  As discussed in
 Appendix~\ref{app.rotation}, choosing different values for the inclination
 (and hence the splitting) does not affect the mode profile, assuming
 reasonable values of the linewidth.

\end{itemize}

We carried out the global fit using both scenarios discussed in
Section~\ref{sec.ridge.id}.  The fit for Scenario~B is shown as the smooth
curve in Figure~\ref{fig.power.zoom} {and the fitted frequencies are
given in Table~\ref{tab.freq.scenarioB}.  Note that the mixed mode at
446\,\muHz\ was not properly fitted because it lies too far from the ridge
(see the first bullet point above).  To check the agreement with the
results discussed in Section~\ref{sec.clean}, we examined the differences
betweens the frequencies in Tables~\ref{tab.freq.matrix}
and~\ref{tab.freq.scenarioB}.  We found a reduced $\chi^2$ of 0.74, which
indicates good agreement.  A value less than 1 is not surprising given that
both methods were constrained to find modes close to the ridges. }

The fitted linewidths (assumed to be a linear function of frequency, as
described above) gave mode lifetimes of
$1.5 \pm 0.4$\,days at 600\,\muHz\ and
$0.6 \pm 0.3$\,days at 1200\,\muHz.
These agree with the single value of $1.29^{+0.55}_{-0.49}$\,days found
above (Section~\ref{sec.lifetimes}), and indicate that the lifetime
increases towards lower frequencies, as is the case for the Sun {and
for the F-type CoRoT targets HD~49933 \citep{BBC2009} and HD~181420
\citep{BDB2009}.}

We also carried out the global fit using Scenario~A.  We found through
Bayesian model selection that Scenario~A was statistically favored over
Scenario~B {by a factor of 10:1.}  This factor classifies as
``significant'' on the scale of \citeauthor{Jef61} (\citeyear{Jef61}; see
Table~1 of \citealt{Lid2009}).  On the same scale, posterior odds of at
least $\sim$13:1 are required for a classification of ``strong to very
strong'', and ``decisive'' requires at least $\sim$150:1.  In our Bayesian
fit to Procyon, the odds ratio in favor of Scenario~A did not exceed 13:1,
even when different sets of priors were imposed.

{In light of the strong arguments given in Section~\ref{sec.ridge.id} in
favour of Scenario~B, we do not consider the result from Bayesian model
selection to be sufficiently compelling to} cause us to reverse our
identification.  {Of course, it is possible that Scenario~A is correct
and, for completeness, we show these fitted frequencies in
Table~\ref{tab.freq.scenarioA}.  The fit using Scenario~A gave mode
lifetimes of
$0.9 \pm 0.2$\,days at 600\,\muHz\ and
$1.0 \pm 0.3$\,days at 1200\,\muHz.}

\section{{Preliminary comparison with} models}

A detailed comparison of the observed frequencies of Procyon with
theoretical models is beyond the scope of this paper, but we will make some
{preliminary comments on the systematic offset between the two.}  It is
well-established that incorrect modeling of the surface layers of the Sun
is responsible for discrepancies between the observed and calculated
oscillation frequencies \citep{ChDDL88,DPV88,RChDN99,LRD2002}.

To address this problem for other stars, \citet{KBChD2008} proposed an
empirical correction to be applied to model frequencies that takes
advantage of the fact that the offset between {observations and models is
independent of $l$} and goes to zero with decreasing frequency.  They
measured the offset for the Sun to be a power law with exponent $b=4.9$ and
applied this correction to the radial modes of other stars, finding very
good results that allowed them to estimate mean stellar densities very
accurately (better than 0.5~per cent).

We have applied this method to Procyon, comparing our observed frequencies
for the radial modes with various published models to determine the scaling
factor $r$ and the offset (see \citealt{KBChD2008} for details of the
method).  The results are shown in Figure~\ref{fig.near.surface}.
Interestingly, the offset between the observations and scaled models does
not go to zero with decreasing frequency.  This contrasts with the G and
K-type stars investigated by \citet{KBChD2008}, namely the Sun, \acena\ and
B, and \bhyi.

The method of \citet{KBChD2008} assumes the correction to be applied to the
models to have the same form as in the Sun, namely a power law with an
exponent of $b=4.9$.  The fit in Figure~\ref{fig.near.surface} is poor and
is not improved by modest adjustments to $b$.  Instead, the results seem to
imply an offset that is constant.  Setting $b=0$ and repeating the
calculations produces the results shown in Figure~\ref{fig.near.surface0},
where {we indeed see a roughly constant offset between the models and the
observations of about 20\,\muHz.}

As a check, we can consider the density implied for Procyon.  The stellar
radius can be calculated from the interferometric radius and the parallax.
The angular diameter of $5.404 \pm 0.031$\,mas \citep[][Table~7]{ALK2005}
and the revised {\em Hipparcos} parallax of $285.93 \pm 0.88$\,mas\
\citep{vanLee2007} give a radius of $2.041 \pm 0.015 \,R_\sun$.

Procyon is in a binary system (the secondary is a white dwarf), allowing
the mass to be determined from astrometry.  \citet{GWL2000} found a value
of $1.497 \pm 0.037\,M_\sun$, while \citet{G+H2006} found $1.431 \pm
0.034\,M_\sun$ {(see \citealt{GKG2008} for further discussion).}

The density obtained using the fits shown in Figure~\ref{fig.near.surface}
is in the range 0.255--0.258\,g\,cm$^{-3}$.  Combining with the radius
implies a mass in the range 1.54--1.56\,$M_\sun$.  The density obtained
using the fits shown in Figure~\ref{fig.near.surface0} is in the range
0.242--0.244\,g\,cm$^{-3}$, implying a mass of 1.46--1.48\,$M_\sun$.  The
latter case seems to be in much better agreement with the astrometrically
determined mass, lending some support to the idea that the offset is
constant.  

We can also consider the possibility that our mode identification is wrong
and that Scenario~A is the correct one (see Sections~\ref{sec.ridge.id}
and~\ref{sec.fit}).  With this reversed identification, the radial modes in
Procyon are those in Table~\ref{tab.freq.matrix} listed as having $l=1$.
Assuming these to be radial modes, the offset between them and the model
frequencies is again constant, as we would expect, but this time with a
mean value close to zero.  {The implied density for Procyon is again
consistent with the observed mass and radius.}

The preceding discussion makes it clear that the correction that needs to
be applied to models of Procyon is very different from that for the Sun and
other cool stars, {regardless of whether Scenario B or A is correct.
In particular, the substantial nearly-constant offset implied by
Figure~\ref{fig.near.surface} would indicate errors in the modeling
extending well beyond the near-surface layers.  We also note that in terms
of the asymptotic expression (equation~\ref{eq.asymptotic}) a constant
offset would imply an error in the calculation of $\epsilon$.}

\section{Conclusion}

We have analyzed results from a multi-site campaign on Procyon that
obtained high-precision velocity observations over more than three weeks
\citep[][Paper~I]{AKB2008PaperI}.  We developed a new method for adjusting
the weights in the time series that allowed us to minimize the sidelobes in
the power spectrum that arise from diurnal gaps and so to construct an
\'echelle diagram that shows two clear ridges of power.  To identify the
odd and even ridges, we summed the power across several orders.  We found
structures characteristic of $l=0$ and 2 in one ridge and $l=1$ and 3 in
the other.  This identification was confirmed by comparing our Procyon data
in a scaled \'echelle diagram \citep{B+K2010} with other stars for which
the ridge identification is known.  We showed that the frequencies of the
ridge centroids and their large and small separations are easily measured
and are useful {diagnostics} for asteroseismology.  In particular, an
oscillation in the large separation appears to indicate a glitch in the
sound-speed profile at an acoustic depth of $\sim$1000\,s.

We identify a strong narrow peak at 446\,\muHz, which falls slightly away
from the $l=1$ ridge, as a mixed mode.  In Table~\ref{tab.freq.matrix} we
give frequencies, extracted using iterative sine-wave fitting, for 55 modes
with angular degrees $l$ of 0--3.  These cover 20 radial orders and a
factor of more than 4 in frequency, which reflects the broad range of
excited modes in Procyon and the high S/N of our data, especially at low
frequencies.  Intensity measurements will suffer from a much higher stellar
background at low frequencies, making it unlikely that even the best data
from the {\em Kepler Mission} will yield the wide range of frequencies
found here.  This is a strong argument in favor of {continuing efforts
towards ground-based Doppler studies, such as} the SONG network (Stellar
Observations Network Group; \citealt{GChDA2008}), {which is currently
under construction, and the proposed Antarctic instrument SIAMOIS (Seismic
Interferometer to Measure Oscillations in the Interior of Stars;
\citealt{MAC2008}). }

We estimated the mean lifetime of the modes by comparing the ``peakiness''
of the power spectrum with simulations and found a value of
$1.29^{+0.55}_{-0.49}$\,days, significantly below that of the Sun.  A
global fit to the power spectrum using Bayesian methods confirmed this
result and provided evidence that the lifetime increases towards lower
frequencies.  {It also casts some doubt on the mode identifications.
We still favor the identification discussed above, but leave open the
possibility that this may need to be reversed.}  Finally, comparing the
observed frequencies of radial modes in Procyon with published theoretical
models showed an offset that {appears to be constant with frequency,
making it very different from that seen in} the Sun and other cool stars.
Detailed comparisons of our results with theoretical models will be carried
out in future papers.

{We would be happy to make the data presented in this paper available on
request.}


\acknowledgments

This work was supported financially by 
the Australian Research Council, 
the Danish Natural Science Research Council,
the Swiss National Science Foundation,
NSF grant AST-9988087 (RPB) and by SUN Microsystems.
{We gratefully acknowledge support from the European Helio- and
Asteroseismology Network (HELAS), a major international collaboration
funded by the European Commission's Sixth Framework Programme.}


\clearpage
\begin{deluxetable}{rrrrr}
\tablecolumns{5}
\tablewidth{0pc}
\tablecaption{Oscillation Frequencies in Procyon (in \muHz) 
\label{tab.freq.matrix}}
\tablehead{
\colhead{Order} & \colhead{$l=0$}  & \colhead{$l=1$}  & \colhead{$l=2$}  & \colhead{$l=3$} }
\startdata
       4 & 
\multicolumn{1}{c}{\nodata} &
  331.3 (0.8) &
\multicolumn{1}{c}{\nodata} &
\multicolumn{1}{c}{\nodata} \\
       5 & 
\multicolumn{1}{c}{\nodata} &
  387.7 (0.7) &
\multicolumn{1}{c}{\nodata} &
\multicolumn{1}{c}{\nodata} \\
       6 & 
  415.5 (0.8) &
  445.8 (0.3) &
  411.7 (0.7) &
\multicolumn{1}{c}{\nodata} \\
       7 & 
  466.5 (1.0) &
  498.6 (0.7) &
  464.5 (0.9) &
  488.7 (0.9) \\
       8 & 
\multicolumn{1}{c}{\nodata} &
  551.5 (0.7) &
\multicolumn{1}{c}{\nodata} &
  544.4 (0.9) \\
       9 & 
  576.0 (0.7) &
  608.2 (0.5) &
\multicolumn{1}{c}{\nodata} &
\multicolumn{1}{c}{\nodata} \\
      10 & 
  630.7 (0.6) &
  660.6 (0.7) &
  627.0 (1.1) &
  653.6 (0.8) \\
      11 & 
  685.6 (0.7) &
  712.1 (0.5) &
  681.9 (0.7) &
\multicolumn{1}{c}{\nodata} \\
      12 & 
  739.2 (0.7) &
  766.5 (0.5) &
  736.2 (0.5) &
\multicolumn{1}{c}{\nodata} \\
      13 & 
  793.7 (0.9) &
  817.2 (0.6) &
  792.3 (0.9) &
\multicolumn{1}{c}{\nodata} \\
      14 & 
  849.1 (0.7) &
  873.5 (0.6) &
  845.4 (0.6) &
  869.5 (0.6) \\
      15 & 
  901.9 (0.8) &
  929.2 (0.7) &
\multicolumn{1}{c}{\nodata} &
  926.6 (0.6) \\
      16 & 
  957.8 (0.6) &
  985.3 (0.7) &
  956.4 (0.5) &
  980.4 (0.9) \\
      17 & 
 1015.8 (0.6) &
 1040.0 (0.7) &
\multicolumn{1}{c}{\nodata} &
 1034.5 (0.7) \\
      18 & 
 1073.9 (0.7) &
 1096.5 (0.7) &
 1068.5 (0.7) &
\multicolumn{1}{c}{\nodata} \\
      19 & 
 1126.7 (0.5) &
 1154.6 (0.9) &
 1124.3 (0.9) &
\multicolumn{1}{c}{\nodata} \\
      20 & 
 1182.0 (0.7) &
 1208.5 (0.6) &
 1179.9 (1.0) &
\multicolumn{1}{c}{\nodata} \\
      21 & 
 1238.3 (0.9) &
 1264.6 (1.0) &
 1237.0 (0.8) &
\multicolumn{1}{c}{\nodata} \\
      22 & 
 1295.2 (1.0) &
\multicolumn{1}{c}{\nodata} &
 1292.8 (1.0) &
\multicolumn{1}{c}{\nodata} \\
      23 & 
 1352.6 (1.1) &
 1375.7 (1.0) &
 1348.2 (1.0) &
\multicolumn{1}{c}{\nodata} \\

\enddata
\end{deluxetable}

\begin{deluxetable}{rr}
\tablecolumns{2}
\tablewidth{0pc}
\tablecaption{Unidentified Peaks with $S/N\ge3.5$ \label{tab.freq.list.other}}
\tablehead{
\colhead{$\nu$}  & \colhead{$S/N$} \\
\colhead{(\muHz)} & \colhead{}  }
\startdata
  407.6 (0.8) & 3.5 \\
  512.8 (0.8) & 3.6 \\
  622.8 (0.6) & 4.3 \\
  679.1 (0.7) & 4.0 \\
  723.5 (0.6) & 4.7 \\
  770.5 (0.7) & 4.1 \\
  878.5 (0.6) & 4.4 \\
  890.8 (0.7) & 3.6 \\
  935.6 (0.7) & 3.9 \\
 1057.2 (0.7) & 3.7 \\
 1384.3 (0.7) & 3.6 \\
\enddata
\end{deluxetable}

\begin{deluxetable}{rrrr}
\tablecolumns{4}
\tablewidth{0pc}
\tablecaption{Frequencies from global fit using Scenario B (in \muHz, with
  $-$/$+$ uncertainties) 
\label{tab.freq.scenarioB}}
\tablehead{
\colhead{Order} & \colhead{$l=0$}  & \colhead{$l=1$}  & \colhead{$l=2$}  }
\startdata
      5 & 
  363.6 (0.8/0.9) &
  387.5 (0.6/0.6) &
  358.5 (1.3/1.2) \\
      6 & 
  415.3 (3.3/1.0) &
\multicolumn{1}{c}{\nodata} &
  408.1 (1.0/3.7) \\
      7 & 
  469.7 (1.6/2.1) &
  498.8 (0.7/0.8) &
  465.3 (1.1/1.3) \\
      8 & 
  522.3 (1.4/1.4) &
  551.6 (0.8/0.7) &
  519.0 (1.5/1.6) \\
      9 & 
  577.0 (1.6/2.5) &
  607.6 (0.6/0.7) &
  573.9 (2.2/2.8) \\
     10 & 
  631.3 (0.8/0.8) &
  660.3 (1.0/1.3) &
  627.4 (2.1/2.8) \\
     11 & 
  685.6 (1.2/1.6) &
  714.7 (1.4/1.2) &
  681.2 (2.3/1.9) \\
     12 & 
  740.1 (1.6/1.7) &
  768.6 (0.9/1.0) &
  737.0 (1.5/1.7) \\
     13 & 
  793.2 (1.3/1.7) &
  820.0 (1.7/1.2) &
  790.9 (2.0/1.9) \\
     14 & 
  847.3 (1.2/1.4) &
  872.7 (1.1/0.9) &
  844.7 (1.7/1.5) \\
     15 & 
  901.0 (1.8/1.7) &
  927.5 (0.8/0.8) &
  898.6 (2.1/2.1) \\
     16 & 
  958.7 (1.4/1.1) &
  983.9 (1.0/1.3) &
  957.2 (1.0/1.3) \\
     17 & 
 1015.9 (1.5/1.8) &
 1039.5 (1.6/1.7) &
 1014.0 (1.8/2.4) \\
     18 & 
 1073.2 (1.5/2.2) &
 1096.6 (1.1/1.0) &
 1070.3 (2.2/2.3) \\
     19 & 
 1127.2 (1.0/1.3) &
 1151.8 (1.4/1.4) &
 1125.9 (1.3/1.4) \\
     20 & 
 1182.3 (1.5/1.4) &
 1207.9 (1.4/1.1) &
 1180.5 (1.6/1.6) \\
     21 & 
 1236.9 (1.7/1.6) &
 1267.4 (1.7/1.5) &
 1235.5 (2.0/1.7) \\
\enddata
\end{deluxetable}

\begin{deluxetable}{rrrr}
\tablecolumns{4}
\tablewidth{0pc}
\tablecaption{Frequencies from global fit using Scenario A (in \muHz, with
  $-$/$+$ uncertainties)
\label{tab.freq.scenarioA}}
\tablehead{
\colhead{Order} & \colhead{$l=0$}  & \colhead{$l=1$}  & \colhead{$l=2$}  }
\startdata
      5 & 
  387.7 (1.9/1.8) &
  361.9 (1.8/2.0) &
  385.1 (1.9/2.6) \\
      6 & 
\multicolumn{1}{c}{\nodata} &
  412.5 (1.7/2.3) &
  439.3 (2.6/2.6) \\
      7 & 
  498.7 (1.1/1.6) &
  467.6 (1.4/1.3) &
  493.2 (2.6/2.0) \\
      8 & 
  552.2 (1.5/1.5) &
  520.7 (1.2/1.3) &
  549.3 (2.2/2.0) \\
      9 & 
  607.8 (1.0/0.9) &
  576.2 (1.1/1.4) &
  605.4 (2.2/2.3) \\
     10 & 
  661.3 (1.3/1.5) &
  631.1 (0.7/0.8) &
  657.1 (1.7/1.6) \\
     11 & 
  716.8 (1.3/1.7) &
  684.7 (1.2/1.2) &
  712.6 (1.2/1.2) \\
     12 & 
  769.9 (1.2/1.3) &
  739.1 (1.1/1.2) &
  766.6 (1.4/1.4) \\
     13 & 
  822.7 (1.9/2.7) &
  792.9 (1.3/1.3) &
  817.8 (1.3/1.4) \\
     14 & 
  874.5 (1.3/1.3) &
  846.4 (0.9/0.8) &
  869.9 (1.6/1.3) \\
     15 & 
  928.8 (1.2/1.2) &
  900.0 (1.3/1.4) &
  925.9 (1.3/1.1) \\
     16 & 
  985.1 (1.0/1.1) &
  958.2 (0.8/0.8) &
  980.9 (1.9/1.6) \\
     17 & 
 1043.4 (2.8/2.8) &
 1015.7 (1.0/0.9) &
 1035.2 (1.0/0.8) \\
     18 & 
 1097.6 (1.5/0.9) &
 1072.5 (1.1/1.2) &
 1091.8 (3.7/4.2) \\
     19 & 
 1153.7 (0.9/0.8) &
 1126.9 (0.5/0.6) &
 1146.8 (1.3/1.0) \\
     20 & 
 1209.1 (0.8/0.9) &
 1181.8 (1.0/0.9) &
 1204.8 (1.3/1.4) \\
     21 & 
 1269.2 (1.0/1.1) &
 1237.1 (0.9/0.9) &
 1264.8 (1.5/1.5) \\
\enddata
\end{deluxetable}


\begin{figure*}
\epsscale{1.0}
\plotone{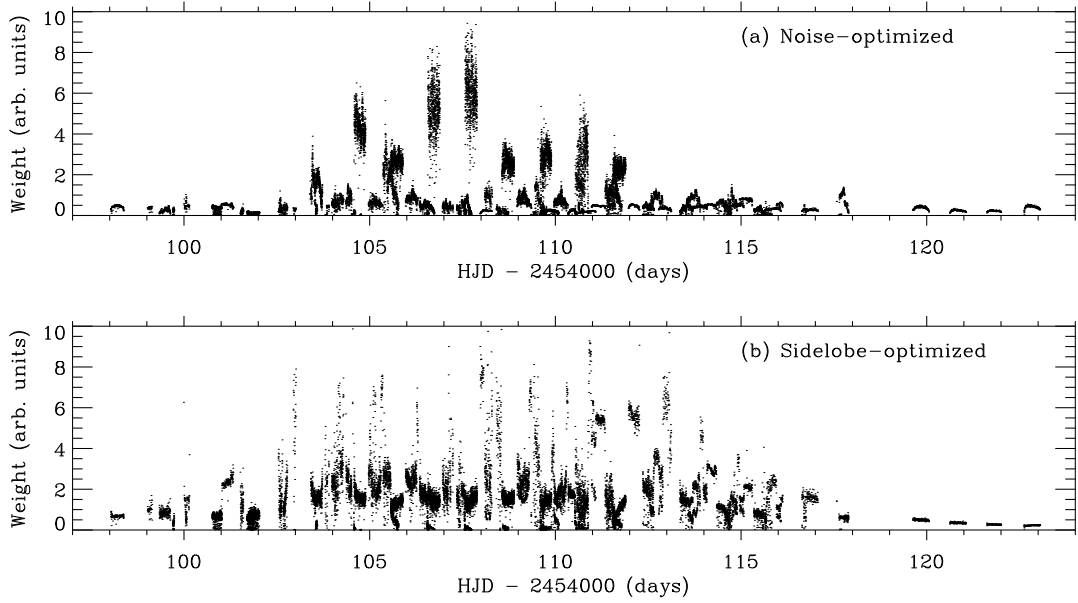}
\caption[]{\label{fig.weights} Weights for time series of velocity
  observations of Procyon, optimized to minimize: ({\em a})~the noise level
  and ({\em b})~the height of the sidelobes.  }
\end{figure*}

\begin{figure*}
\epsscale{1.0}
\plotone{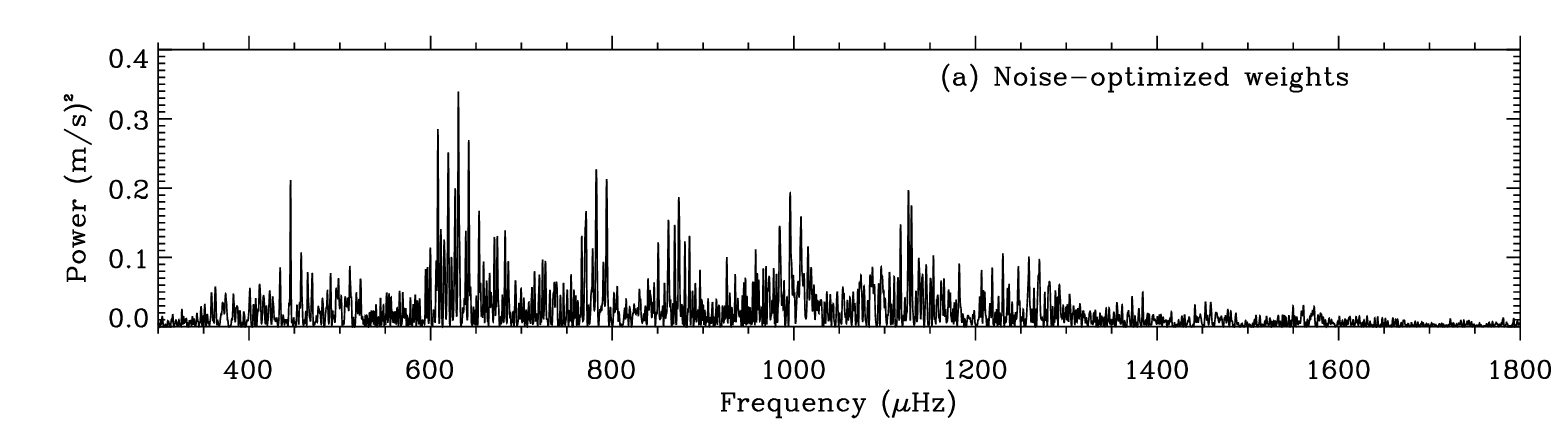}
\plotone{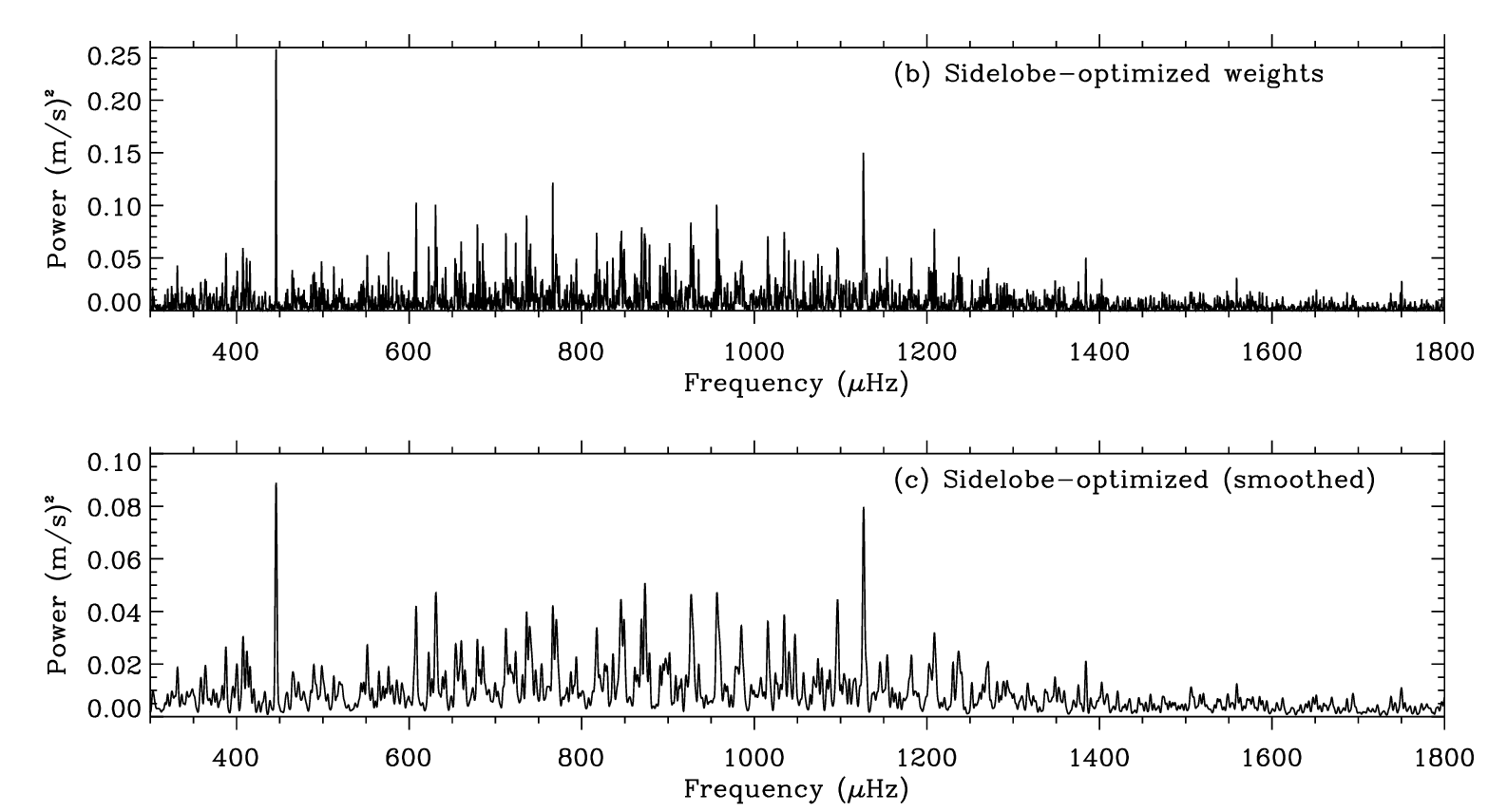}
\caption[]{\label{fig.power} Power spectrum of oscillations in Procyon:
 ({\em a})~using the noise-optimized weights;
 ({\em b})~using the sidelobe-optimized weights;
 ({\em c})~using the sidelobe-optimized weights and smoothing
  by convolution with a Gaussian with FWHM 2\,\muHz. 
}
\end{figure*}

\clearpage

\begin{figure}
\epsscale{1.2}
\plotone{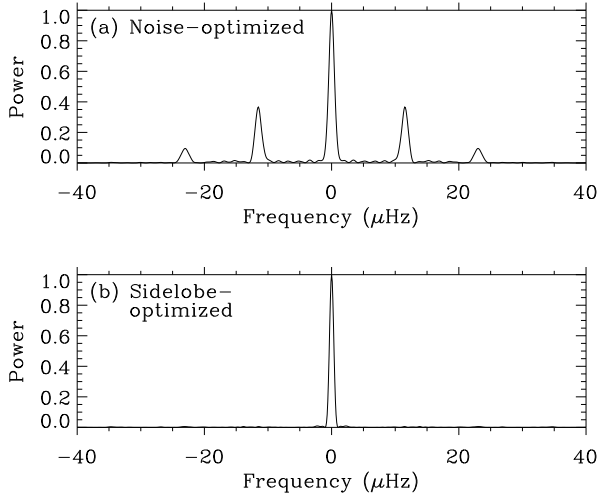}
\caption[]{\label{fig.window} Spectral window for the Procyon observations
  using ({\em a})~noise-optimized weights and ({\em b})~sidelobe-optimized
  weights. }
\end{figure}

\begin{figure}
\epsscale{1.2}
\plotone{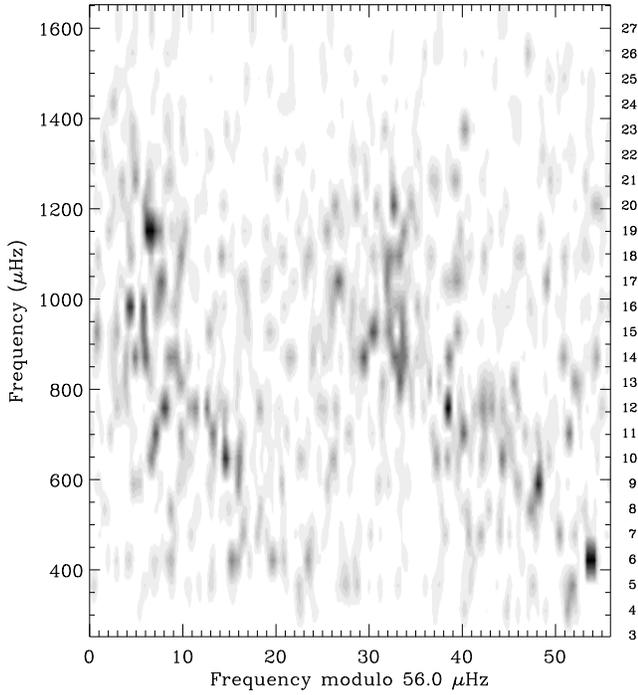}
\caption[]{\label{fig.echelle.image.WIN56} Power spectrum of Procyon in
echelle format using a large separation of 56\,\muHz, based on the
sidelobe-optimized weights.  
Two ridges are clearly visible.  The upper parts are vertical
but the lower parts are tilted, indicating a change in the large separation
as a function of frequency.  The orders are numbered sequentially on the
right-hand side.}
\end{figure}

\begin{figure}
\epsscale{1.2}
\plotone{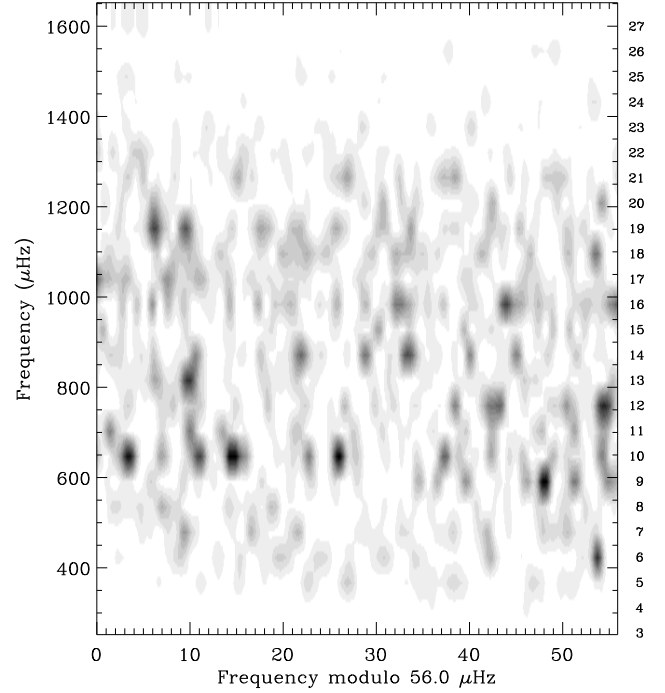}
\caption[]{\label{fig.echelle.image.SNR56} Same as
Fig.~\ref{fig.echelle.image.WIN56}, but for the noise-optimized weights.
The sidelobes from daily aliasing mean that the ridges can no longer be
clearly distinguished. }
\end{figure}

\begin{figure}
\epsscale{1.2}
\plotone{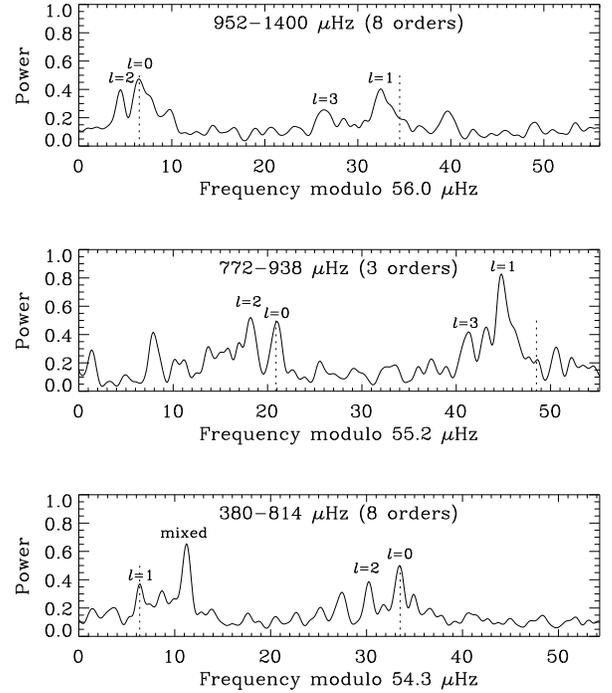}
\caption[]{\label{fig.collapse3}The power spectrum of Procyon collapsed
  along several orders.  Note that the power spectrum was first smoothed
  slightly by convolving with a Gaussian with FWHM 0.5\,\muHz.
  The dotted lines are separated by exactly $\Dnu/2$, to guide the
  eye in assessing the 0--1 small separation}
\end{figure}


\begin{figure}
\epsscale{1.2}
\plotone{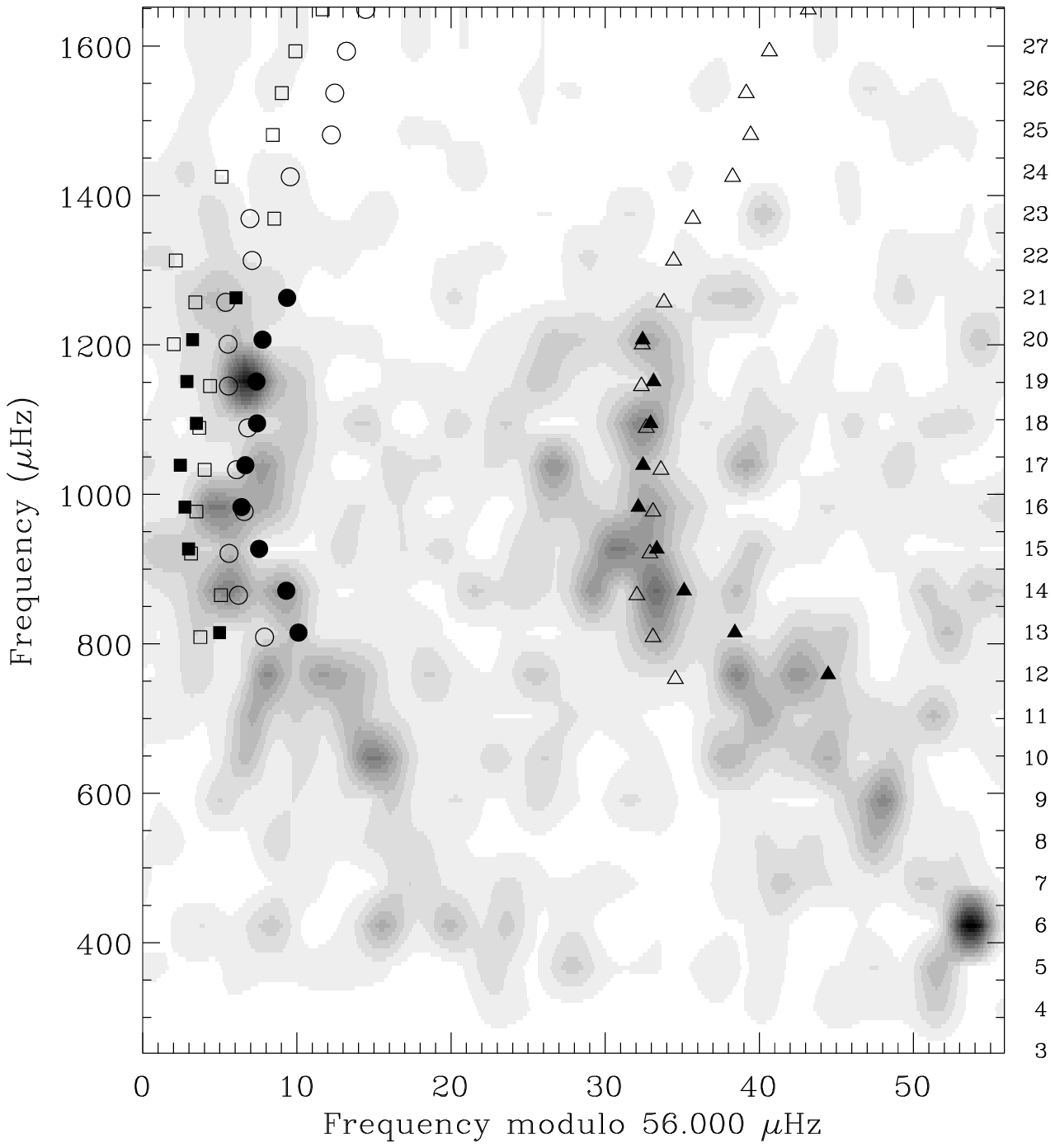}
\caption[]{\label{fig.echelle.corot} \'Echelle diagram for Procyon smoothed
     to 2\,\muHz\ (greyscale) overlaid with scaled frequencies for two
     stars observed by CoRoT.  The filled symbols are oscillation
     frequencies for HD~49385 reported by \citet{DBM2010}, after
     multiplying by 0.993.  Open symbols are oscillation frequencies for
     HD~49933 from the revised identification by
     \citet[][Scenario~B]{BBC2009} after multiplying by 0.6565.  Symbol
     shapes indicate mode degree: $l=0$ (circles), $l=1$ (triangles), and
     $l=2$ (squares).}
\end{figure}


\begin{figure}
\epsscale{1.2}
\plotone{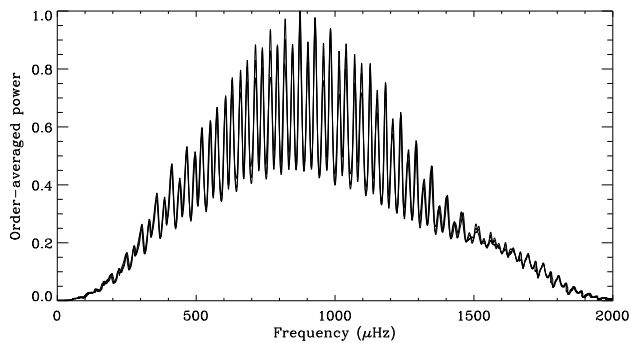}
\caption[]{\label{fig.idl9} Order-averaged power spectrum (OAPS), where
  smoothing was done with a FWHM of 4.0 orders (see text).  The OAPS is
  plotted for three values of the large separations (54, 55 and 56\,\muHz)
  and we see that the positions of the maxima are not very sensitive to the
  value of \Dnu.}
\end{figure}


\begin{figure}
\epsscale{1.2}
\plotone{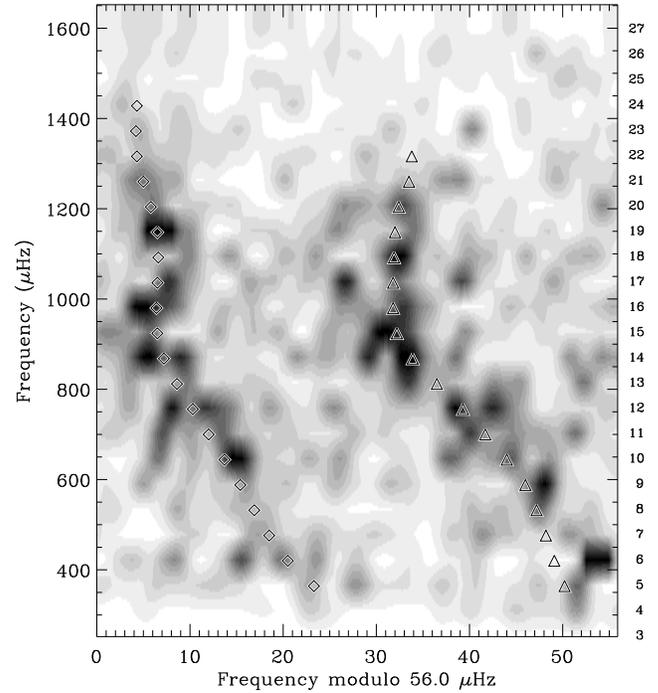}
\caption[]{\label{fig.echelle.ridges56} Centroids of the two ridges, as
measured from the comb response.  The grayscale shows the
sidelobe-optimized power spectrum from which the peaks were calculated.}
\end{figure}


\begin{figure}
\epsscale{1.2}
\plotone{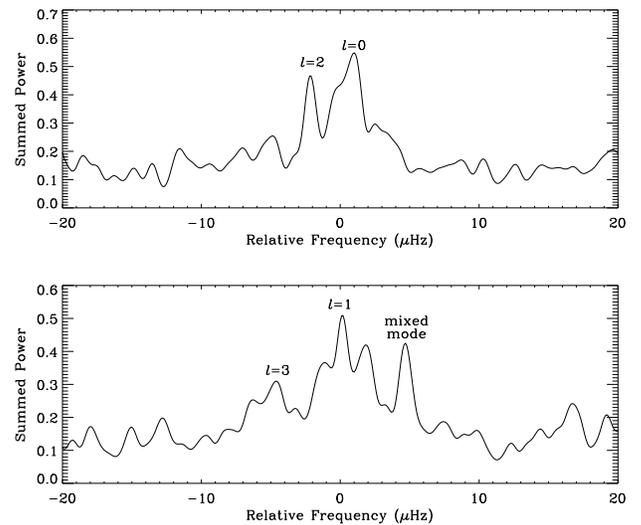}
\caption[]{\label{fig.collapse.ridges}The power spectrum of Procyon
  collapsed along the ridges, over the full range of oscillations (18
  orders).  {The upper panel shows the left-hand ridge, which we
  identify with modes having even degree, and the lower panel shows the
  right-hand ridge (odd degree).}  Note that the power spectrum was first
  smoothed slightly by convolving with a Gaussian with FWHM 0.6\,\muHz. }
\end{figure}


\begin{figure}
\epsscale{0.6}
\plotone{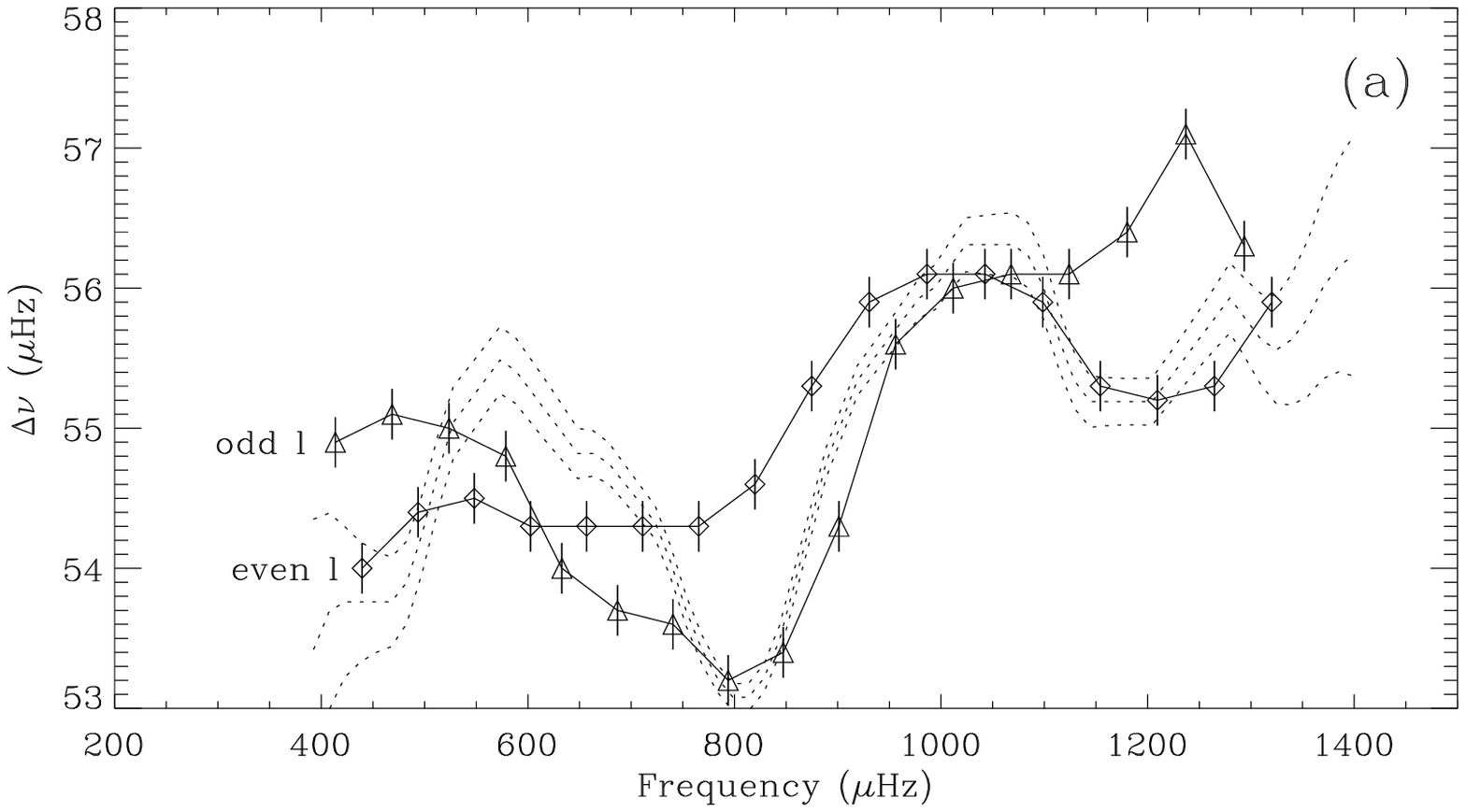}
\plotone{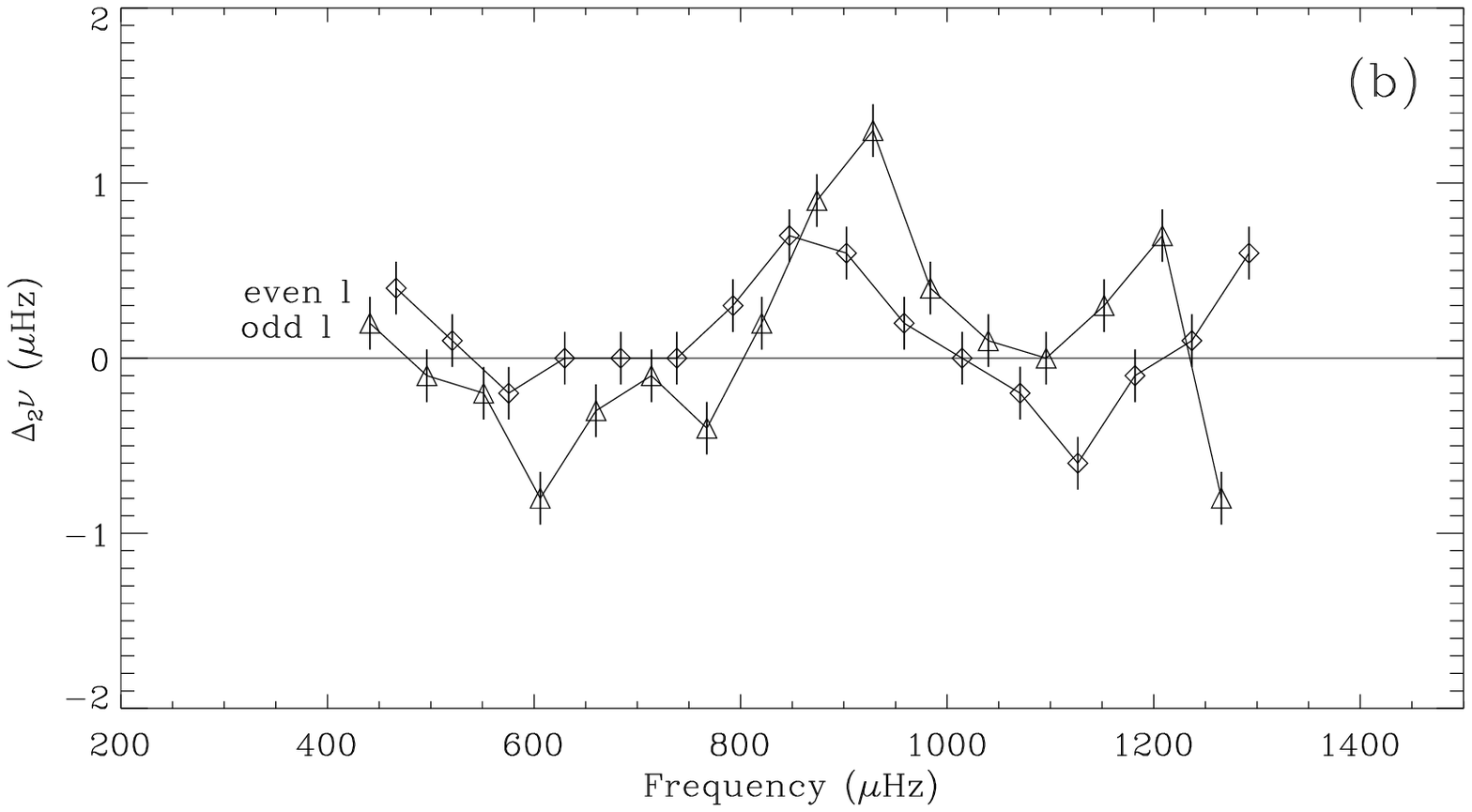}
\plotone{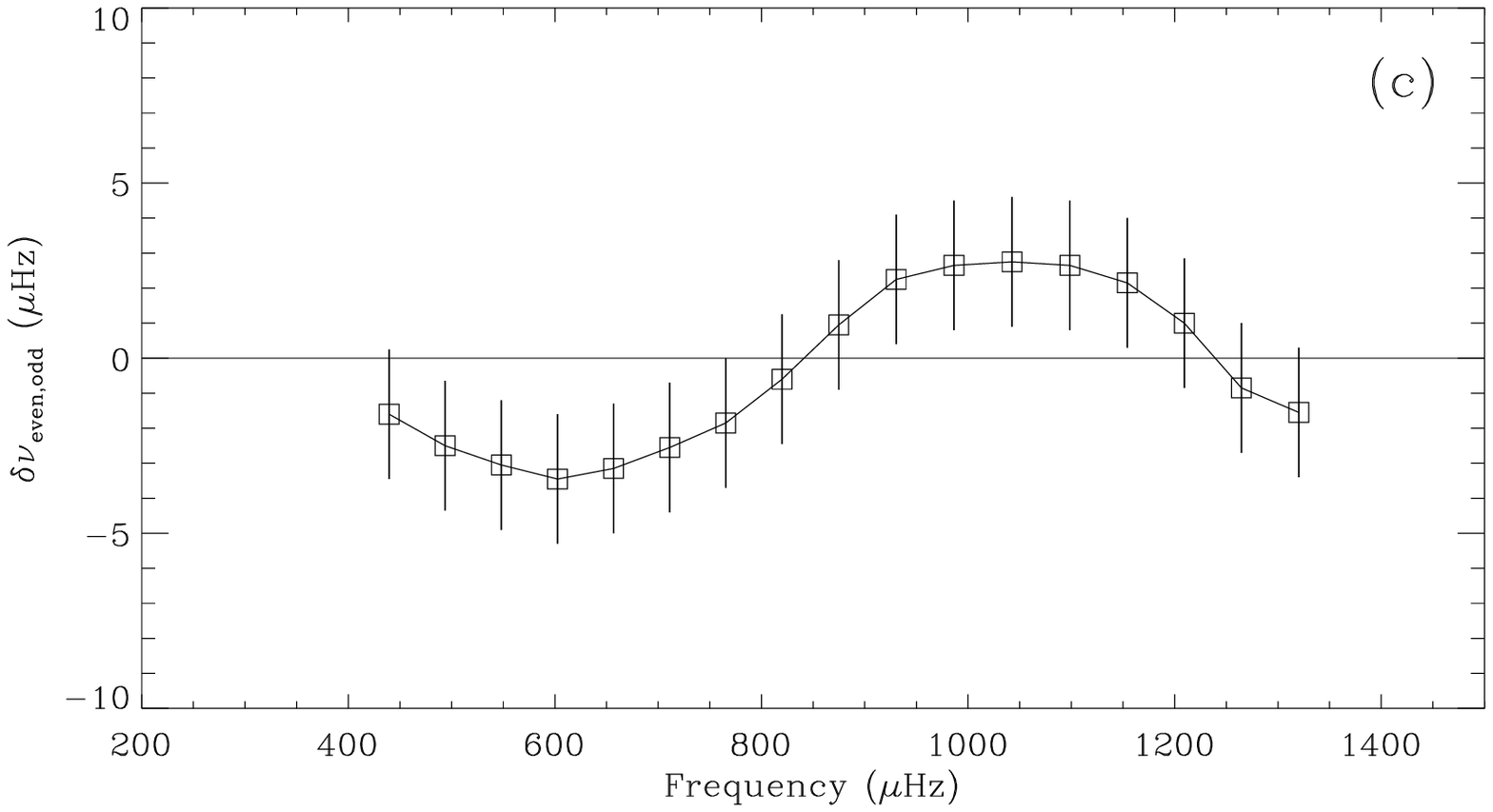}
\caption[]{\label{fig.seps.ridges} Symbols show the frequency separations
  in Procyon as a function of frequency, as measured from the ridge
  centroids: (a) large frequency separation, (b) second differences, and
  (c) small frequency separation.  The dotted lines in panel~{\it a} show
  the variation in \Dnu\ (with $\pm1\sigma$ range) calculated from the
  autocorrelation of the time series -- see the text.} 
\end{figure}


\begin{figure}
\epsscale{1.2}
\plotone{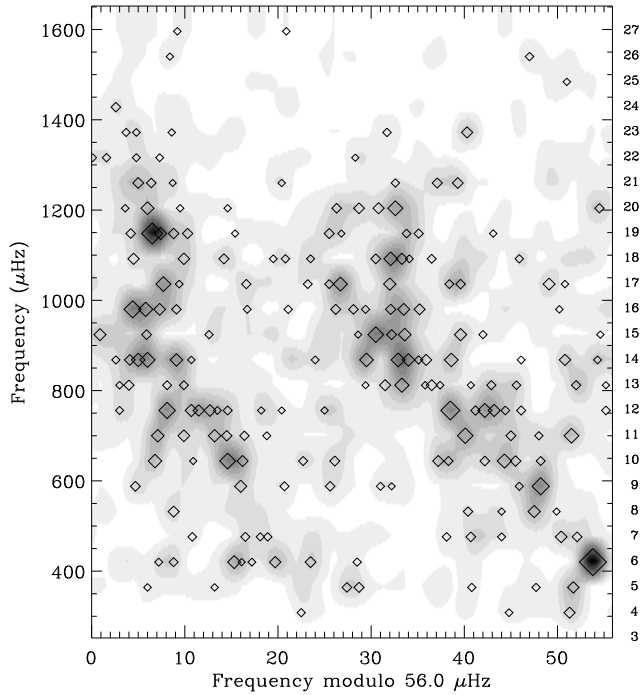} 
\caption[]{\label{fig.echelle.clean.sidelobe} Peaks extracted from
  sidelobe-optimized power spectrum using iterative sine-wave fitting.
  Symbol size is proportional to amplitude (after the background noise has
  been subtracted).  The grayscale shows the sidelobe-optimized power
  spectrum on which the fitting was performed, to guide the eye.}
\end{figure}

\begin{figure*}
\epsscale{1.0}
\plotone{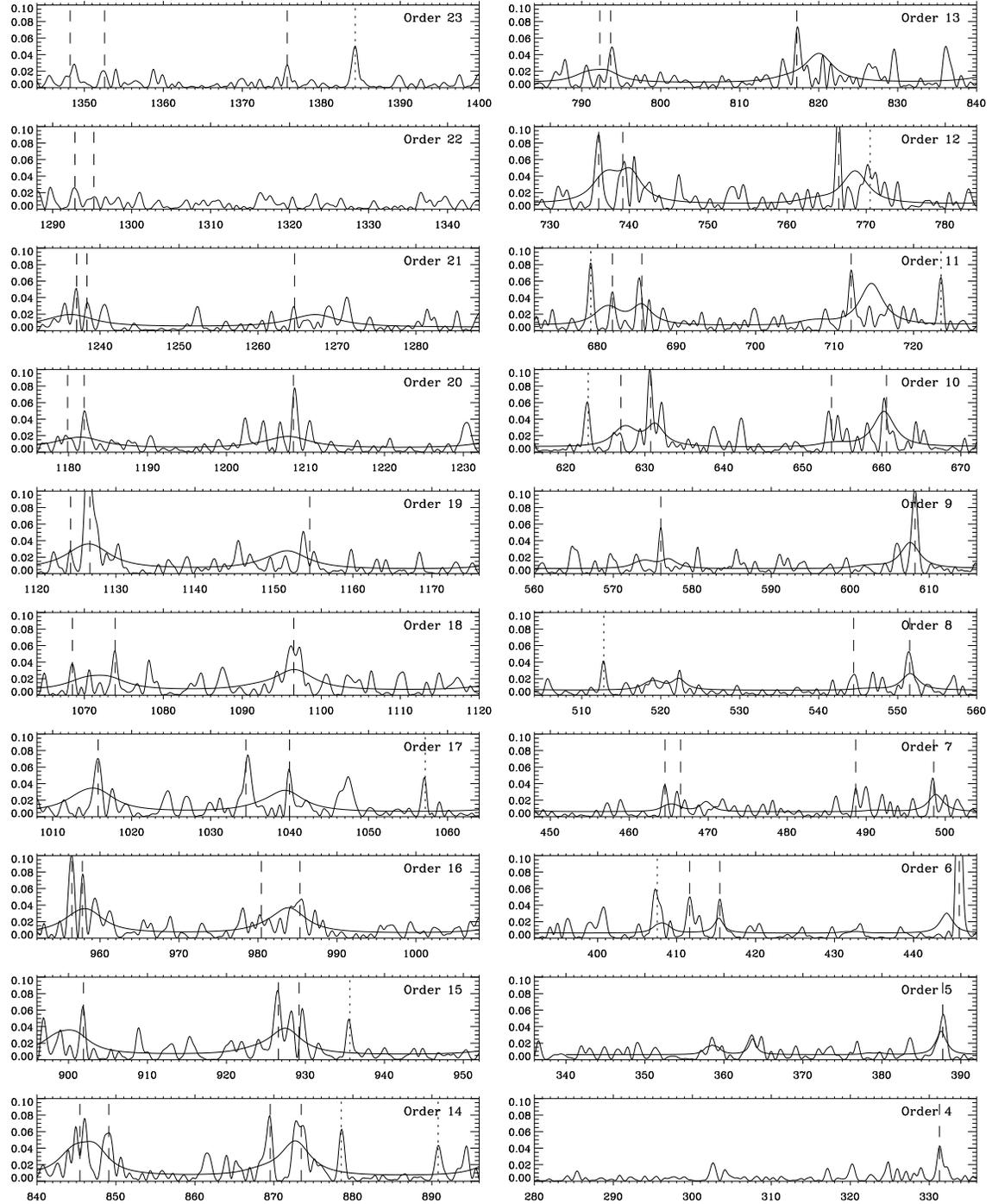}  
\caption[]{\label{fig.power.zoom}The power spectrum of Procyon at full
  resolution, {with the orders in each column arranged from top to
  bottom, for easy comparison with the \'echelle diagrams.}  Vertical dashed
  lines show the mode frequencies listed in Table~\ref{tab.freq.matrix} and
  dotted lines show the peaks that have not been identified, as listed in
  Table~\ref{tab.freq.list.other}.  The smooth curve shows the global fit
  to the power spectrum for Scenario~B (see Section~\ref{sec.fit}). }
\end{figure*}

\begin{figure}
\epsscale{1.2}
\plotone{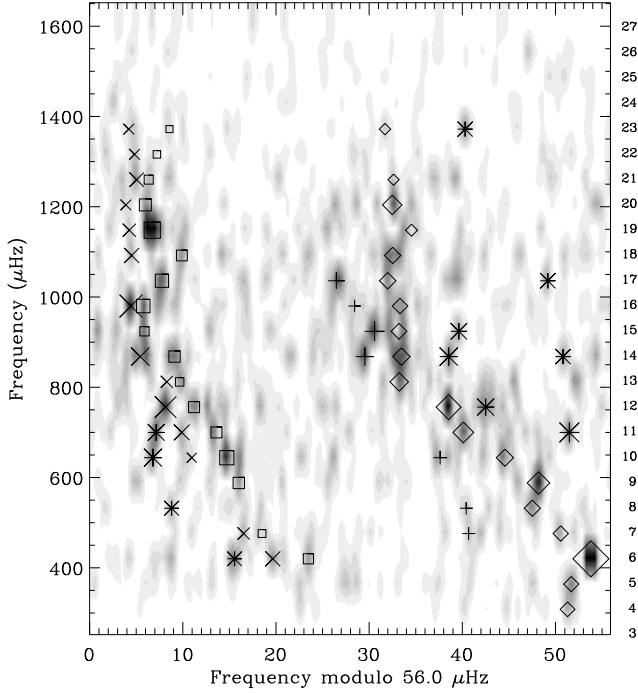}  
\caption[]{\label{fig.echelle.cleanid} The power spectrum of Procyon
  overlaid with mode frequencies listed in Table~\ref{tab.freq.matrix}.
  Symbols indicate angular degree (squares: $l=0$; diamonds: $l=1$;
  crosses: $l=2$; pluses: $l=3$).  Asterisks show the peaks that have not
  been identified, as listed in Table~\ref{tab.freq.list.other}.  }
\end{figure}

\begin{figure}
\epsscale{1.2}
\plotone{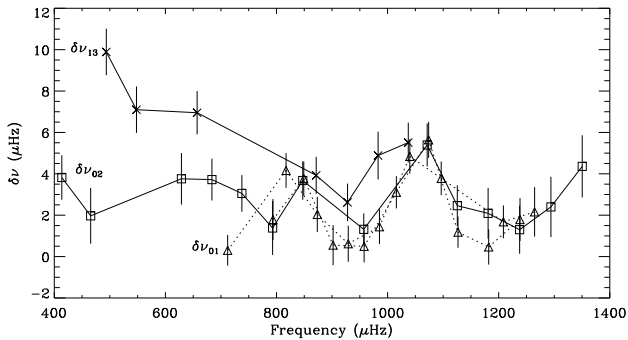}  
\caption[]{\label{fig.seps.freq} Small frequency separations in Procyon, as
  measured from the mode frequencies listed in
  Table~\ref{tab.freq.matrix}. }
\end{figure}

\begin{figure}
\epsscale{1.2}
\plotone{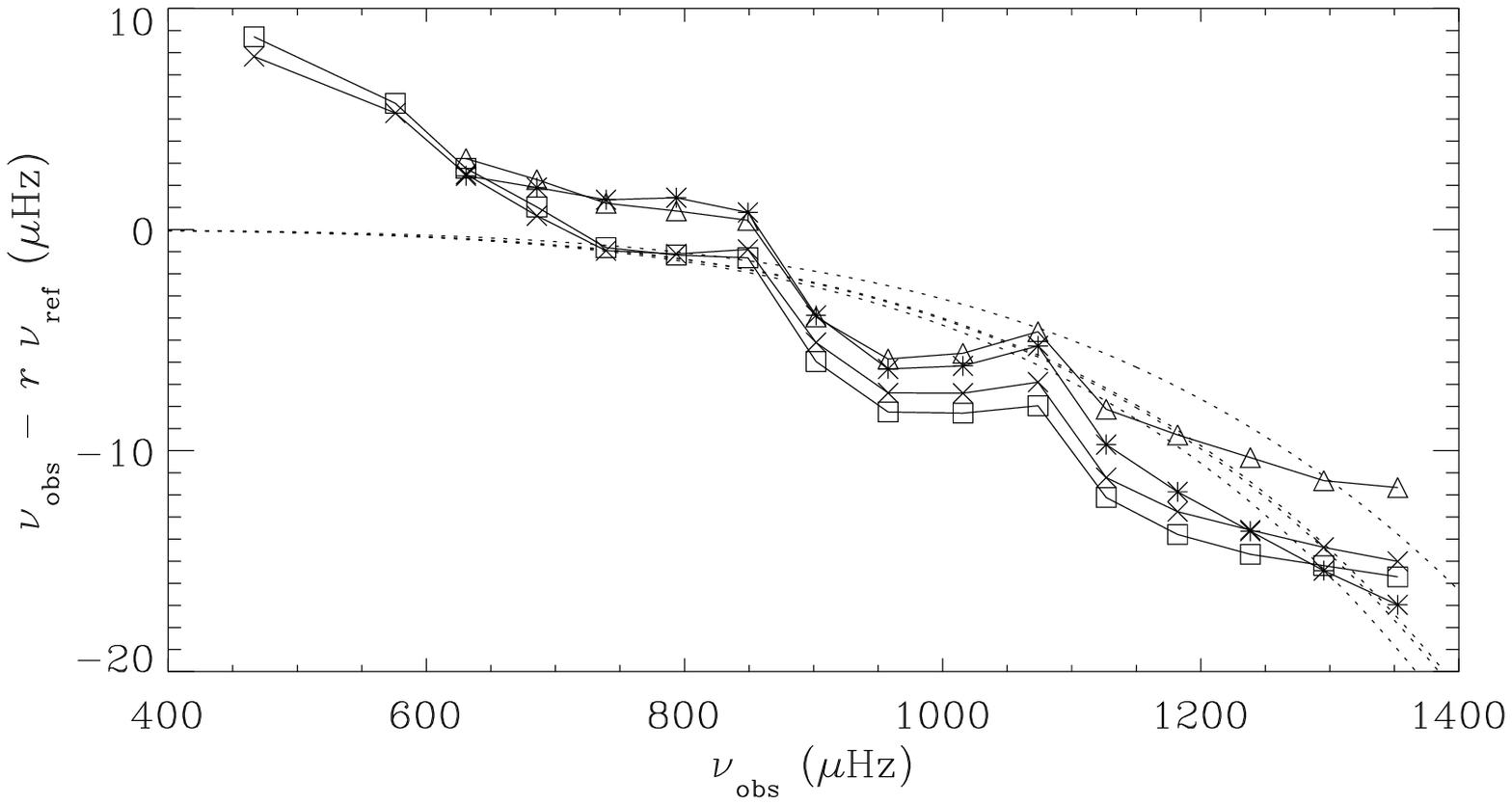}  
\caption[]{\label{fig.near.surface} The difference between observed
frequencies of radial modes in Procyon and those of scaled models.  The
symbols indicate different models, as follows: 
squares from \citet[][Table~2]{CDG99}, 
crosses from \citet[][]{DiM+ChD2001}, 
asterisks from \citet[][Table~4]{KTM2004}, and
triangles from \citet[][model M1a]{ECB2005}.
In each case, the dotted curve shows the correction calculated using
equation~(4) of \citet{KBChD2008}.}
\end{figure}

\begin{figure}
\epsscale{1.2}
\plotone{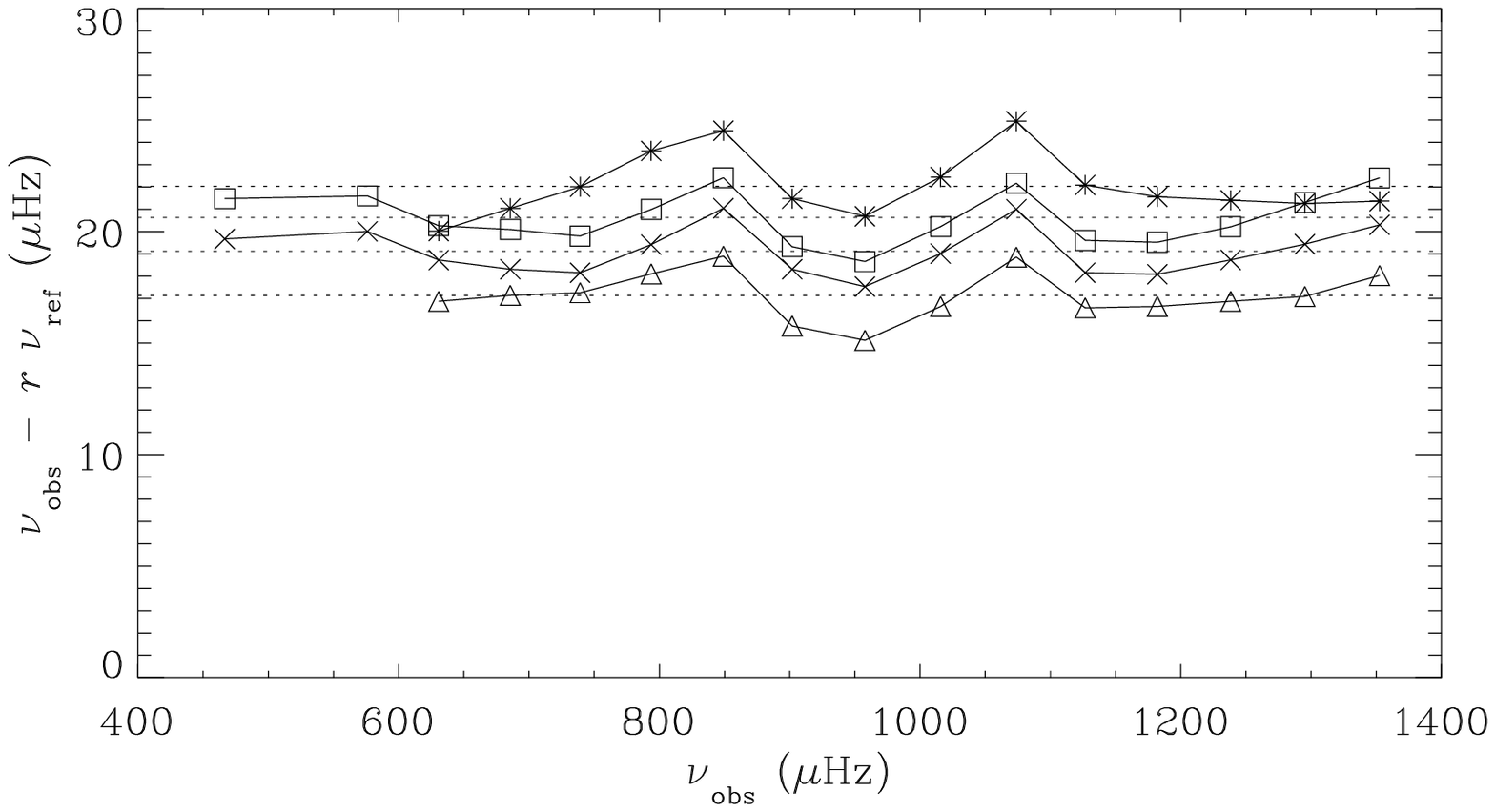}  
\caption[]{\label{fig.near.surface0} Same as
 Figure~\ref{fig.near.surface}, but with a constant near-surface
 correction ($b=0$). }
\end{figure}

\appendix

\section{Rotational splitting}\label{app.rotation}

We expect non-radial modes to be split due to the rotation of the star.
The rotation period of Procyon is not known, although slow variations in
our velocity observations (Paper~I) indicated a value of either $10.3$\,days
or twice that value.  The projected rotational velocity has been measured
spectroscopically.  \citet{APAL2002} determined a value of $v\sin i = 3.16
\pm 0.50$\,\kms{}, although they note that the actual value may be lower by
about 0.5\,\kms.

\citet{G+S2003} have studied the effect of rotation on the profiles of
solar-like oscillations as a function of inclination and mode lifetime
\citep[see also][]{BGL2006}.  We have repeated their calculations for our
observations of Procyon (with sidelobe-optimized weights).  The results are
shown in Figure~\ref{fig.rotation}, which shows the effects of rotational
splitting, inclination angle and mode lifetime on the theoretical profile
of the modes.\footnote{Note that we have made the quite reasonable
assumption that the internal rotation has a similar period to the surface
rotation.}  Note that the calculations do not include the stochastic nature
of the excitation and so the function shown here should properly be called
the expectation value of the power spectrum, also known as the {\em limit
spectrum}.  Figure~\ref{fig.rotation} is similar to Figure~2 of
\citet{G+S2003} except that instead of fixing the rotation period, we have
fixed $v\sin i$ to be the measured value.  For $l=0$ the profile does not
depend on the inclination angle, while for $l=1$, 2 and~3 the solid and
dashed lines show calculations for $i=30^{\circ}$ ($P_{\rm rot} =
16.4$\,days) and $i=80^{\circ}$ ($P_{\rm rot} = 32.3$\,days), respectively.
In each panel, results are shown for three values of the mode lifetime:
1.5\,days (top), 3\,days (middle) and infinite (bottom).  For each mode
lifetime, the curves for different $i$ and $l$ are all normalized to have
the same area.

We see from Figure~\ref{fig.rotation} that for a fixed $v\sin i$, the width
of the profile stays roughly constant as a function of inclination.  If the
rotation axis of the star happens to be in the plane of the sky
($i=90^{\circ}$) then the rotation period is too low to produce a
measurable splitting.  At the other extreme, if the inclination is small
(so that the rotation is close to pole-on), then the rotational splitting
will be large but most of the power will be in the central peak ($m=0$).
Either way, once the profile has been broadened by the mode lifetime, the
splitting will be unobservable.

We conclude that for realistic values of the mode lifetime, our
observations are not long enough to detect rotational splitting in Procyon.
The line profiles are broadened by rotation, but it is not possible to
disentangle the rotation rate from the inclination angle.  Rotational
splitting is not measurable in Procyon, except perhaps with an extremely
long data set.  The detection of rotational splitting requires choosing a
star with a larger $v\sin i$ or a longer mode lifetime, or both.

\section{Relating ridge centroids to mode  frequencies}
\label{app.ridges}

As discussed in Section~\ref{sec.ridge.centroids}, the frequencies of the
ridge centroids are useful for asteroseismology in cases where it is
difficult to resolve the ridges into their component modes.  In this
appendix, we relate the frequencies of the ridge centroids to those of the
underlying modes, which allows us to express the small separation of the
ridges (equation~\ref{eq.dnu_even_odd}) in terms of the conventional small
separations (\dnu{01}, \dnu{02}, and \dnu{13}).  These relationships will
allow the observations to be compared with theoretical models.

The ridge centroids depend on the relative contributions of modes with
$l=0$, 1, 2, and 3.  The power in the even ridge is approximately equally
divided between $l=0$ and $l=2$, while the odd ridge is dominated by $l=1$
but with some contribution from $l=3$.  The exact ratios depend on the
observing method, as discussed by \citet{KBA2008}.  For velocity
measurements, such as those presented in this paper for Procyon, the
amplitude ratios given by \citet[][their Table~1]{KBA2008} yield the
following expressions for the centroids in power:
\begin{eqnarray}
  \nu^{\rm vel}_{n, \rm even} &=& 0.49 \nu_{n,0} + 0.51 \nu_{n-1,2} \\
  \nu^{\rm vel}_{n, \rm odd}  &=& 0.89 \nu_{n,1} + 0.11 \nu_{n-1,3},
\end{eqnarray}
where the superscript indicates these apply to velocity measurements.

For photometric measurements, such as those currently being obtained with
the CoRoT and Kepler Missions, the relative contributions from the various
$l$ values are different.  Table~1 of \citet{KBA2008} gives response
factors for intensity measurements in the three VIRGO passbands, namely
402, 500 and 862\,nm.  For CoRoT and Kepler, it is appropriate to use a
central wavelength of 650\,nm.  Using the same method as \citet{KBA2008},
we find the ratios (in amplitude) for this case to be $S_0:S_1:S_2:S_3 =
1.00:1.23:0.71:0.14$.  The ridge centroids measured from such data would
then be
\begin{eqnarray}
  \nu^{650}_{n, \rm even} &=& 0.66 \nu_{n,0} + 0.34 \nu_{n-1,2} \\
  \nu^{650}_{n, \rm odd}  &=& 0.99 \nu_{n,1} + 0.01 \nu_{n-1,3},
\end{eqnarray}

We can express the new small separation of the ridge centroids
(equation~\ref{eq.dnu_even_odd}) in terms of the conventional ones.  For
velocity we have
\begin{equation}
  \delta\nu^{\rm vel}_{\rm even,odd} = \delta\nu_{01} - 0.51 \delta\nu_{02} + 0.11 \delta\nu_{13}.
\end{equation}
and for photometry we have
\begin{equation}
  \delta\nu^{650}_{\rm even,odd} = \delta\nu_{01} - 0.34 \delta\nu_{02} + 0.01 \delta\nu_{13}.
\end{equation}

Finally, we can express these in terms of $D_0$ under the assumption that
the asymptotic relation (equation~\ref{eq.asymptotic}) holds exactly,
although in fact this is not likely to be the case:
\begin{equation}
  \delta\nu^{\rm vel}_{\rm even,odd} = 0.04 D_0
\end{equation}
and
\begin{equation}
  \delta\nu^{650}_{\rm even,odd} = 0.06 D_0.
\end{equation}

\begin{figure}
\epsscale{1.2}
\epsscale{0.7}
\plotone{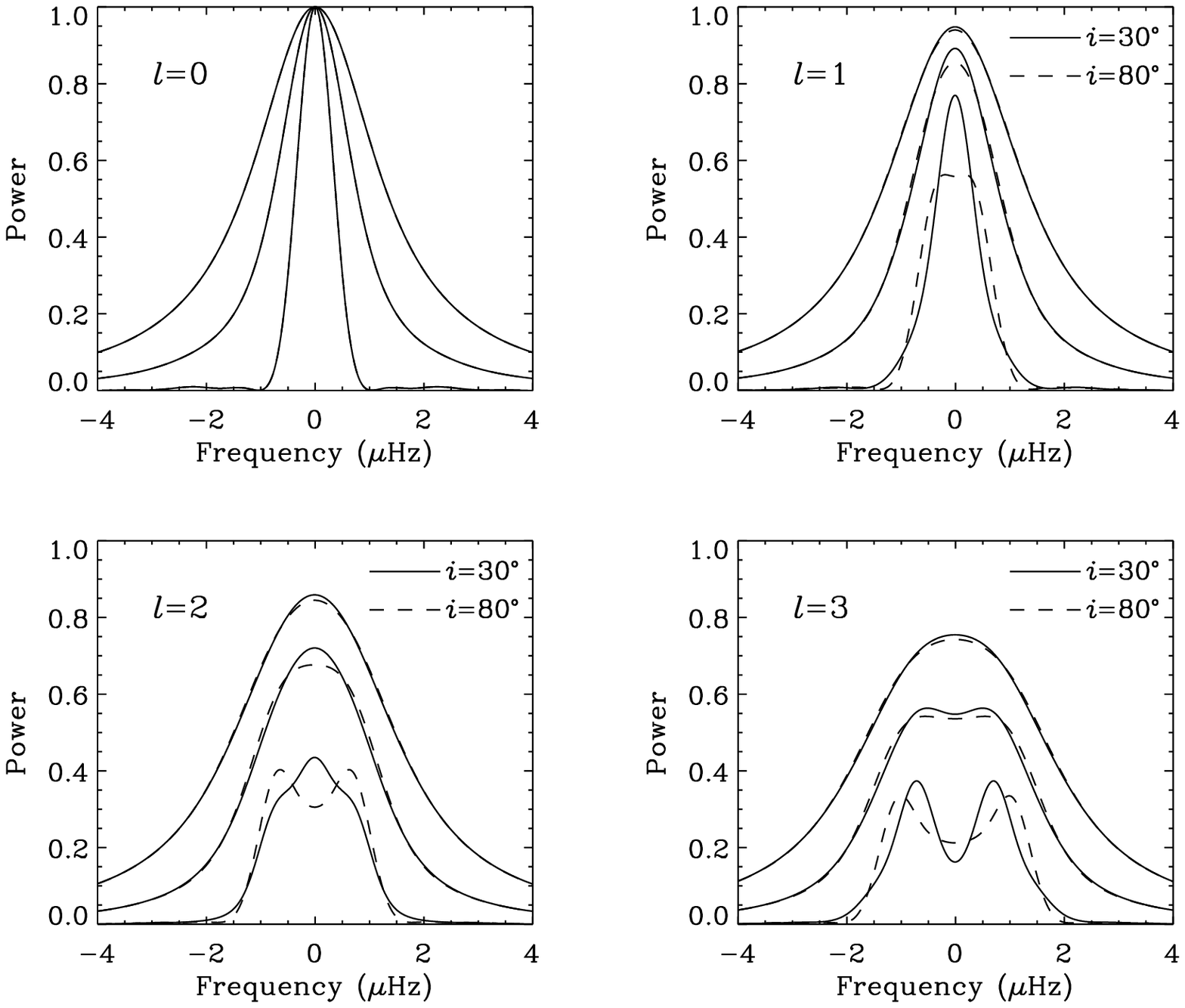}  
\caption[]{\label{fig.rotation} Theoretical line profiles showing
  rotational splitting for different mode degrees, similar to Fig.~2 of
  \citet{G+S2003} but here using a fixed value of $v\sin i$, namely
  3.16\,\kms, as measured for Procyon \citep{APAL2002}).  For $l=0$ the
  profile does not depend on the inclination angle, while for $l=1$, 2
  and~3 the solid and dashed lines show calculations for $i=30^{\circ}$
  ($P_{\rm rot} = 16.4$\,days) and $i=80^{\circ}$ ($P_{\rm rot} =
  32.3$\,days), respectively.  In each panel, results are shown for three
  values of the mode lifetime: 1.5\,days (top), 3\,days (middle) and
  infinite (bottom).  For each mode lifetime, the curves for different $i$
  and $l$ are all normalized to have the same area. }
\end{figure}

\typeout{get arXiv to do 4 passes: Label(s) may have changed. Rerun}

\end{document}